\documentclass{article}
\pdfoutput=1
\usepackage{amsmath}
\usepackage[utf8]{inputenc}
\usepackage{natbib}
\usepackage{appendix}
\usepackage{graphicx}
\usepackage{subcaption}
\usepackage{parskip}
\usepackage{hyperref}
\usepackage{soul}
\usepackage[affil-it]{authblk}

\hypersetup{
    colorlinks=true,
    linkcolor=blue,
    filecolor=blue,      
    urlcolor=blue,
    citecolor = blue,
    pdftitle={AD Study},
    pdfpagemode=FullScreen,
}

\usepackage{tikz}
\usetikzlibrary{decorations.pathreplacing}
\usetikzlibrary{shapes.geometric}
\usetikzlibrary{shapes.multipart}
\usetikzlibrary{arrows,topaths}

\newcommand{\boldtheta}{\mbox{\boldmath{$\theta$}}}
\newcommand{\boldTheta}{\mbox{\boldmath{$\Theta$}}}
\newcommand{\boldpsi}{\mbox{\boldmath{$\psi$}}}
\newcommand{\boldlambda}{\mbox{\boldmath{$\lambda$}}}

\newcommand{\boldbeta}{\mbox{\boldmath{$\beta$}}}

\newcommand{\boldeta}{\mbox{\boldmath{$\eta$}}}

\newcommand{\boldnu}{\mbox{\boldmath{$\nu$}}}
\newcommand{\boldOmega}{\mbox{\boldmath{$\Omega$}}}

\newcommand{\boldomega}{\mbox{\boldmath{$\omega$}}}

\newcommand{\boldgamma}{\mbox{\boldmath{$\gamma$}}}
\newcommand{\smallboldgamma}{\mbox{\tiny \boldmath{$\gamma$}}}

\newcommand{\boldPhi}{\mbox{\boldmath{$\Phi$}}}

\newcommand{\normal}{\mbox{N}}
\newcommand{\by}{\mathbf{y}}

\newcommand{\bX}{\mathbf{X}}

\newcommand{\bx}{\mathbf{x}}
\newcommand{\bV}{\mathbf{V}}
\newcommand{\bv}{\mathbf{v}}

\newcommand{\bz}{\mathbf{z}}

\newcommand{\bm}{\mathbf{m}}

\newcommand{\bZ}{\mathbf{Z}}

\newcommand{\bW}{\mathbf{W}}
\newcommand{\bw}{\mathbf{w}}
\newcommand{\bM}{\mathbf{M}}

\newcommand{\bR}{\mathbf{R}}

\title{Bayesian Variable Selection in Distributed Lag Models: A Focus on Binary Quantile and Count Data Regressions}

\author[1,*]{Daniel Dempsey}
\author[1]{Jason Wyse}

\date{April 2024}

\affil[1]{School of Computer Science and Statistics, Trinity College Dublin}
\affil[*]{Corresponding author email: daniel.dempsey0@gmail.com}

\begin{document}

\maketitle

\begin{abstract}
\noindent Distributed Lag Models (DLMs) and similar regression approaches such as MIDAS have been used for many decades in econometrics and more recently to investigate how poor air quality adversely affects human health. In this paper we describe how to expand the utility of these models for Bayesian inference by leveraging latent variables. In particular we explain how to perform binary regression to better handle imbalanced data, how to incorporate negative binomial regression, and how to estimate the probability of predictor inclusion. Extra parameters introduced through the DLM framework may require calibration for the MCMC algorithm, but this will not be the case in DLM-based analyses often seen in pollution exposure literature. In these cases, the parameters are inferred through a fully automatic Gibbs sampling procedure.
\end{abstract}

\emph{Keywords:} Distributed lag regression, P\'olya-Gamma regression, Functional data analysis, Air quality, Pollution exposure, Environmental modelling.

\section{Introduction} \label{sec:intro}
It can often be the case that variables are correlated in a manner that accumulates over time. For example, the extent to which exposure to poor air quality affects human health will depend on both the concentration of poor quality air and how long ago the exposure occurred. It is reasonable to assume that the effect of exposures will asymptotically decline over time, but there is no guarantee that the decline will be monotonic in the short term. Uncovering the underlying temporal behaviour is key to understanding the effects of these exposures as a whole. \emph{Distributed Lag Models} (DLMs) are specifically built to properly incorporate these longitudinal predictors into a regression model. 

DLMs first gained popularity within econometrics \citep{almon1965distributed, hannan1965estimation}, and in recent times it has been adopted in healthcare research \citep{schwartz2000, zanobetti2002temporal, warren2020spatially}. A related class of models known as \emph{Mixed Data Sampling} (MIDAS) models was introduced by \cite{ghysels2004midas} to analyse data with mixed sampling regimes (for example, measuring the association of a time series with observations sampled yearly and another with observations sampled monthly). The utility of DLMs (and MIDAS) is evident from the number of analyses centred around them; \cite{wilson2017bayesian} used a Bayesian DLM to investigate the relationship between air quality and adverse health outcomes of newborns. \cite{warren2020spatially} implemented random effects to account for spatial autocorrelation in their DLM when studying the effect of air quality exposure during pregnancy and the birth weight of newborns. \cite{antonelli2021multiple} developed a Bayesian DLM that utilises spike-and-slab priors for variable selection \citep{mitchell1988bayesian}. \cite{mogliani2020} similarly suggest a MIDAS model with spike-and-slab priors, but with a Laplacian distribution for the slab to incorporate Bayesian Group-LASSO variable selection \citep{meier2008group, xu2015bayesian}. In this article, we will also be discussing variable selection, adapting the method presented by \cite{holmes2006} into the DLM framework. The advantage of this method over the sparsity approaches is that it allows us to directly estimate the probability of predictor inclusion. Depending on how the DLM is specified, it is possible to accomplish this without dimensional transitions in the MCMC routine.

Our focus in this manuscript is placed entirely on binary and count response data. We will use quantile methods to fit binary response models \citep{benoit2017} since the skewed link function may better handle imbalanced binary response data \citep{czado1992effect}, and count data will be handled using negative binomial regression \citep{pillow2012fully, zhou2012lognormal}. We will show how both of these methods allow for easy incorporation of the \cite{holmes2006} variable selection algorithm.

The remainder of this article is laid out as follows: in Section~\ref{sec:method} we present the general framework for generalised DLMs, focusing on how we can apply binary quantile regression and negative binomial regression in particular. We outline a simulation study in Section~\ref{sec:sim_study} for assessing the accuracy of the methods at estimating associated parameters and how well the variable selection procedure performs. There is an additional document supplementary to this manuscript that contains even more simulation study results. In Section~\ref{sec:data} the proposed approach is demonstrated on a real data example. Both the real and simulated data examples are based on Chicago air quality data available in the \texttt{dlnm} package for \texttt{R} \citep{gasparrini2011distributed}. We conclude in Section~\ref{sec:conclusion}.

\section{Methods} \label{sec:method}
Our goal is to fit a regression model that explains the association between a response variable $\by$ and a collection of predictors, some of which have a standard contemporaneous effect (referred to as \emph{static} predictors) and others that have a longitudinal effect (referred to as \emph{dynamic} predictors). We will denote the matrix of static predictors (which includes the intercept, if used) as $\bZ_\text{S}$. 

The dynamic predictors will be represented by matrices containing their lags. For ease of notation and explanation we will assume we have only one dynamic predictor, and thus only one corresponding lag matrix, denoted $\bZ$. Each row of this matrix contains the lags of the dynamic variable from the corresponding time index as follows:

\begin{equation} \label{eq:lag_mat}
    \bZ = 
    \begin{bmatrix}
    z_T & z_{T-1} & \dots & z_{T-\tau} \\
    z_{T-1} & z_{T-2} & \dots & z_{T-1-\tau} \\
    \vdots & \vdots & & \vdots \\
    z_{T-n} & z_{T-n-1} & \dots & z_{T-n-\tau} 
    \end{bmatrix}
\end{equation}

where $z_t$ is the value of the dynamic variable observed at time $t$, $T$ is the most recently observed time index, $n$ is the number of response observations, and $\tau$ defines how many lags are related to the response, set by the user. Without loss of generality, \eqref{eq:lag_mat} assumes that the contemporaneously observed lag affects the value of the response but, of course, this is not necessary and can be changed to start at whichever lag the analyst deems appropriate.

With this, DLMs can be expressed in the following general form:

\begin{align}
    &\label{eq:base_dlm} \by \sim F\left( g^{-1}(\boldeta), \boldPhi \right)\\
    &\label{eq:eta} \boldeta = \bZ_\text{S}\boldbeta_\text{S} + \bZ\bW(\boldTheta)
\end{align}

\noindent where $F$ is the assumed distribution of the response $\by$, $g^{-1}(\cdot)$ is the inverse link function, and $\boldPhi$ is the collection of parameters of $F$ that are required along with the transformed linear predictor.

$\bW(\cdot)$ encodes the relationship between the lags of the dynamic predictor and the response via some operation over the lag indices, which will usually (though not necessarily) depend on some collection of parameters denoted as $\boldTheta$. We will refer to $\bW(\cdot)$ as the \emph{lag-response} throughout this paper, following the nomenclature established in \cite{gasparrini2013modeling}. 

The exact form of the lag-response is determined by how the analyst chooses to model the relationship between the response and the lags. For example, one could choose to use a linear combination of the all the lags, effectively treating each lag as its own predictor \citep{schwartz2000}. In this case, the lag-response is the identity matrix multiplied by a vector of linear coefficients. This is straightforward to fit as it is effectively a simple GLM, but such models often exhibit identification issues \citep{foroni2015unrestricted}. 

In MIDAS literature, it is common to model the lag-response as a weighted aggregate of the lags multiplied by some magnitude parameter. The weights are determined by a fitted curve called the \emph{distributed lag function}, or DLF \citep{lutkepohl1981model, ghysels2007midas}. In this case $\bW(\boldTheta) = \bw(\boldtheta)\beta$ where $\bw(\boldtheta)$ is a vector of weights, adjusted according to the chosen DLF and $\boldtheta$ parameters. A list of typical choices of DLFs is given in Table 1 of \cite{ghysels2016}. While the DLF approach offers an appealing compromise between model flexibility and analyst control, the non-linear relationship between $\by$ and $\boldtheta$ potentially adds a large challenge to model fitting.

Another option - and the one we will be adopting in this manuscript - is to use a linear combination of a basis expansion of the lag indices \citep{gasparrini2010distributed}. In this case, the lag-response is a basis matrix multiplied by a vector of linear coefficients; the specific method we will be using throughout this paper is a B-spline expansion. The major consideration for spline fitting are the placement of knots. A common heuristic is to simply place equally spaced knots across the data (in our case, the lag indices), though more sophisticated approaches can be used, such as using penalisation to achieve a desired level of smoothness \citep{eilers2010splines, rushworth2013distributed} or using reversible-jump MCMC to fit the knots \citep{dimatteo2001bayesian}. Beyond these concerns around the knots, this approach is simply another linear combination and thus \eqref{eq:base_dlm} and \eqref{eq:eta} collapses to a generalised linear model.

The drawback with the basis expansion approach is that, unlike DLFs, it is harder to constrain the curves such that unreasonable models are completely excluded from the posterior. For example, it is possible to end up with a lag-response that swaps sign within the time window, thus resulting in, for example, a model that claims a pollutant can be both detrimental \emph{and} beneficial depending on how long ago the exposure occurred. This may be a reasonable trade-off if such models are given negligible weight in the posterior.

\subsection{Model Inference}
Going back to \eqref{eq:base_dlm}, in this manuscript we are primarily concerned with the situation where $F$ is a discrete distribution, used to model either count or binary responses. When estimating the posterior, straightforward application of Bayes' Theorem in either case leads to full conditionals that cannot be sampled directly regardless of the choice of prior. These are recognised as challenging modelling problems for Bayesian methods. However, in both cases we can alleviate this issue with the introduction of latent variables. First let us introduce some extra notation; let $\bX$ be the column-wise concatenation of $\bZ_\text{S}$ and $\bZ\bW(\boldTheta)$, and let $\boldbeta$ (with the subscript omitted), denote the vector containing $\boldbeta_S$ and the subset of all $\boldTheta$ that are linear coefficients. Note that when the lag-response represents a linear combination of basis expansions then $\boldbeta$ encompasses all the model's parameters.

\subsubsection{Count Data Inference}
Automatic Gibbs sampling for the parameters of the negative binomial regression model is explained by \cite{pillow2012fully} and \cite{zhou2012lognormal}, which are briefly summarised here. We start by assuming $y_t$ (in this case, a discrete count) follows a negative binomial distribution with probability parameter equal to the transformed linear predictor and unknown stopping parameter $\xi$,

\begin{eqnarray*}
    y_t &\sim& \text{NB}( \xi, p_t )\\
    p_t &=& \frac{\exp(\eta_t)}{1 + \exp(\eta_t)}
\end{eqnarray*}

\noindent where $\eta_t$ are the components of $\boldeta$ given in \eqref{eq:eta}. Assuming a Gaussian prior of mean vector $\bm$ and covariance matrix $\bv$, exact and automatic Gibbs sampling from the posterior of $\boldbeta$ can be accomplished by introducing a latent variable $\boldomega$ that follows a P\'olya-Gamma distribution \citep{polson2013},

\begin{eqnarray*}
    \omega_t \,|\, \boldbeta, \xi &\sim& \text{PG}\left( y_t + \xi,\, \bz^\top_{\text{S},t}\boldbeta_\text{S} + \bz_t^\top\bW(\boldTheta) \right)\\
    \boldbeta \,|\, \boldomega, \xi &\sim& \normal\left(\bM_C, \bV_C\right)\\
    && \\
    \bV_C &=& \left(\bX^\top\boldOmega_C\bX + \bv^{-1}\right)^{-1}\\
    \bM_C &=& \bV_C\left(\bX^\top\boldOmega_C\boldlambda_C + \bv^{-1}\bm\right)\\
    \boldOmega_C &=& \text{diag}(\boldomega)\\
    \lambda_{t,C} &=& (y_t - \xi)/2\omega_t
\end{eqnarray*}

\noindent where diag($\cdot$) constructs a diagonal matrix based on the components of the vector input. The subscript $C$ is used throughout to highlight that these are the parameters corresponding to the count data model; this is to avoid notational conflict in the next section. Regarding the P\'olya-Gamma distribution, Section 4 of \cite{polson2013} describes an exact and efficient rejection sampler which they have implemented in their \texttt{R} package \texttt{BayesLogit}.

To sample from the posterior of $\xi$, we set a gamma distribution prior with shape parameter $a_0$ and scale parameter $b_0$. By introducing another latent variable $\boldpsi$, $\xi$ is updated as follows:

\begin{equation*}
	\pi(\psi_t = j\,|\,\xi) = R_{y_t, j} \nonumber
\end{equation*}
\begin{equation*}
    \xi \sim \text{Gamma}\left( a_0 + \sum_{i=1}^N \psi_i, b_0 - \sum_{i=1}^N\ln(1 - p_t) \right) \nonumber 
\end{equation*}

\noindent where $N$ is the length of $\by$ and $R_{i, j}$ is the $i, j\text{th}$ element of the following (recursively constructed) lower triangular matrix: 

\begin{equation} \label{eq:xi_mat}
	R_{i, j} = \frac{i - 1}{i}R_{i - 1, j} + \frac{\xi}{i} R_{i - 1, j - 1}
\end{equation}

\noindent with $R_{1, 1} = R_{0, 1} = 1$, and $R_{i, j} = 0$ when $j = 0$ or $j > i$. This matrix has square dimensions equal to the maximum value of $\by$. Construction of this matrix is the biggest potential source of computational bottleneck as it has to be re-calculated every iteration, cell-by-cell. An alternative non-recursive method exists \citep{zhou2012lognormal} but it quickly becomes numerically unstable as either $\xi$ or the dimension of the $\bR$ matrix grow too large.

Poisson regression can be approximated by simply fixing $\xi$ to some large number but otherwise following the above approach, circumventing the need to compute \eqref{eq:xi_mat}. See \cite{dangelo2022efficient} for further notes on approximate Bayesian Poisson regression. \cite{zhou2012lognormal} augmented the above algorithm to capture broader forms of variation, and also derived associated Variational Bayes estimates.

\subsubsection{Possibly Imbalanced Binary Data Inference}
Now let us assume $y_t$ is binary. Exact Gibbs sampling for probit regression (and an approximation for logistic regression) via latent variables was first proposed by \cite{albert1993} and has become the blueprint for binary response regression; see \cite{holmes2006, fruhwirth2010data, polson2013, zens2023ultimate} for more examples of how it is applied for logistic regression. Here, we wish to use this same approach as a way of handling DLMs for binary responses, where the response may be imbalanced. This can occur in, say, medical studies where the focus is on rare diseases and/or symptoms. 

The MLE of logit regression parameters has been shown to be biased when the logit misspecifies the `true' link, and that this bias is diminished by skewness \citep{czado1992effect}. For this reason, the approach we recommend here is binary quantile regression \citep{benoit2017} using the 3-parameter \emph{Asymmetric Laplace Distribution} (ALD) \citep{yu2005three}, with skew parameter $q \in (0, 1)$ that corresponds to the desired quantile. Define $\boldbeta$ as in the previous section. The ALD Bayesian quantile model is

\begin{align*}
    &y_t = I(y_t^\star > 0)\\
    &y_t^\star = \bz_{\text{S},t}^\top\boldbeta_\text{S} + \bz_t^\top\bW(\boldTheta) + \epsilon_t\\
    &\epsilon_t \sim \text{ALD}(\mu = 0, \, \sigma = 1, \, q)
\end{align*}

\noindent where $I(\cdot)$ is equal to 1 if the condition inside the parentheses is met and 0 otherwise. $\mu$ and $\sigma$ are location and scale parameters for the ALD. \cite{kozumi2011gibbs} found that the ALD can be expressed as location-scale mixture of Gaussians by introducing a standard exponential variable $\boldnu$. The full-conditional distribution of $\boldnu$ is

\begin{equation} \label{eq:nu_fc}
    \nu_t\,|\,y^\star_t, \boldbeta \sim \text{GIG}(1/2, \chi_t^2, \delta^2)
\end{equation}

\noindent where

\[    \chi_t^2 = \frac{(y^\star_t - \bx_t^\top\boldbeta)^2}{\phi^2} \,,\quad    \delta^2 = 2 + \frac{\psi^2}{\phi^2}\,,\quad 
    \psi^2 = \left(\frac{1 - 2q}{q(1 - q)}\right)^2 \,,\quad  
    \phi^2 = \frac{2}{q(1 - q)}.
\]
The full conditional distribution of $\by^\star$ is truncated Gaussian, but as noted in \cite{holmes2006}, we can instead sample from the full conditional \emph{joint} distribution of $\boldnu$ and $\by^\star$ by noting that $\pi(\boldnu, \by^\star| \boldbeta) = \pi(\boldnu|\by^\star, \boldbeta)\pi(\by^\star|\boldbeta)$. The first factor is simply \eqref{eq:nu_fc} as before, but the update to $\by^\star$ is now independent of $\boldnu$, thus improving Monte Carlo efficiency. Its distribution is a truncated ALD, 

$$
y^\star_t | y_t, \boldbeta \sim 
\begin{cases}
	\text{ALD}(\bx_t^\top\boldbeta, \sigma = 1, q) & \text{restricted to the positive axis when $y_t = 1$,}\\
	\text{ALD}(\bx_t^\top\boldbeta, \sigma = 1, q) & \text{restricted to the negative axis when $y_t = 0$}
\end{cases}
$$

\noindent which is easily sampled from using the inversion method, or via the approach given in section 4.2 of \cite{benoit2017}.

As with negative binomial inference, the resulting full conditional distribution of the linear coefficients is Gaussian,

\begin{equation*}
	\boldbeta\,|\,\by, \boldnu \sim \normal(\bM_B, \bV_B)
\end{equation*}
\noindent where
\begin{eqnarray*}
    \bV_B &=& \left(\bX^\top \boldOmega_B \bX + \bv^{-1}\right)^{-1}\\
	\bM_B &=& \bV_B\left(\bX^\top \boldOmega_B\boldlambda_B + \bv^{-1}\bm\right)\\
	\boldOmega_B &=& \left(\phi^2\text{diag}(\boldnu)\right)^{-1}\\
	\boldlambda_B &=& \by^\star - \psi\boldnu
\end{eqnarray*}

\noindent and as before, $\bm$ and $\bv$ are the mean and covariance of the prior Gaussian distribution for $\boldbeta$.

The user must decide the skew parameter (quantile) to use for the ALD link. As suggested by \cite{kordas2006}, estimation can be improved in the presence of rare event data (many zeros) when a larger quantile is selected, and vice-versa.

In the event that the lag-response $\bW(\boldTheta)$ from equation \eqref{eq:eta} is chosen in such a way that all the $\boldTheta$ components are linear coefficients, then this is all that is necessary for fitting DLMs with count or binary responses. As stated before, any non-linear parameters that arise from the lag-response model (as usually seen when applying DLFs) will require further consideration.

\subsection{Covariate Inclusion Inference}
Here we discuss how to adapt the models above to infer predictor inclusion uncertainty, using a method proposed by \cite{holmes2006}. First, let $\gamma_i \in \{0, 1\}$ represent whether or not a corresponding predictor (or groups of predictors) be included in the model. We can update $\boldgamma$ jointly with $\boldbeta$ through a Metropolis update, 

\begin{equation*}
	q(\boldbeta^\star, \boldgamma^\star) = \pi(\boldbeta^\star\, |\, \boldgamma^\star, \mathcal{D})q( \boldgamma^\star\, |\, \boldgamma) 
\end{equation*}

\noindent where $\mathcal{D}$ denotes every parameter in the model (including latent variables). The first component is simply the full conditional distribution of $\boldbeta$ already derived for both models above, but only using the columns of $\bX$ supported by $\boldgamma^\star$. The second is the user-chosen method for proposing new values of $\boldgamma$. Under this proposal scheme, the acceptance rate is

\begin{equation*}
	\alpha = \min\left\{1, \frac{|\bv_{\smallboldgamma^\star}|^{-0.5}|\bV_{\smallboldgamma^\star}|^{0.5}\exp\left[0.5 \left(\bM^\top_{\smallboldgamma^\star}\bV^{-1}_{\smallboldgamma^\star}\bM_{\smallboldgamma^\star} + \bm_{\smallboldgamma^\star}^\top\bv^{-1}_{\smallboldgamma^\star}\bm_{\smallboldgamma^\star} \right)\right]q(\boldgamma\,|\,\boldgamma^\star)\pi_{\smallboldgamma}(\boldgamma^\star)}{|\bv_{\smallboldgamma}|^{-0.5}|\bV_{\smallboldgamma}|^{0.5}\exp\left[0.5 \left(\bM^\top_{\smallboldgamma}\bV^{-1}_{\smallboldgamma}\bM_{\smallboldgamma} + \bm_{\smallboldgamma}^\top\bv^{-1}_{\smallboldgamma}\bm_{\smallboldgamma}\right)\right]q(\boldgamma^\star\,|\,\boldgamma)\pi_{\smallboldgamma}(\boldgamma)}\right\}
\end{equation*}

\noindent where $\pi_{\smallboldgamma}(\cdot)$ is the prior of $\boldgamma$. Going forward, the prior on each $\gamma_i$ is assumed to be Bernoulli with success probability parameter equal to 0.5. Here, $\bM_{\smallboldgamma}$ and $\bV_{\smallboldgamma}$ are the posterior mean and variance of $\boldbeta$ from the appropriate model, calculated using only the predictors permitted by $\boldgamma$. Notice that $\boldbeta$ is not present in the formula, so no further adjustments to the acceptance probability is required. This makes sense as `removing' a variable from the above regression models is the same as simply fixing its associated value at zero, so there is actually no shift in the dimensionality when any element of $\boldgamma$ is altered. This is no longer the case if the lag-response uses non-linear parameters, in which case altering $\boldgamma$ will now result in genuine dimension jumping and the acceptance rate will need to be adjusted accordingly \citep{green1995, green2009}.

In regards to the proposal $q(\boldgamma^\star|\boldgamma)$, the approach we use in this paper is to randomly select a predictor each iteration and propose to swap the current value of the corresponding $\gamma$ value from 0 to 1 or vice-versa. This is a symmetric proposal, so the $q(\cdot)$ terms cancel in the acceptance probability. Note this becomes unsuitable as the number of potential predictors grow larger since the probability that any specific value is swapped at any iteration shrinks to zero. Adapting sparsity methods instead \citep{carvalho2010horshoe, piironen2017sparsity} would be preferable in such situations, though with the drawback that they don't directly compute the probability of inclusion. See also \cite{ji2012} for a similar approach that uses Gibbs sampling to update the inclusion parameters instead of Metropolis-within-Gibbs.

\section{Simulation Study} \label{sec:sim_study}
A simulation study was performed to test the above methods. The simulated data were partially based on a table of air quality metrics collected between 1987 and 2000 in Chicago, available through the \texttt{R} package \texttt{dlnm} \citep{gasparrini2011distributed}. Three columns from this dataset were used as dynamic predictors: rhum (mean relative humidity), PM10 (particulate matter), and O3 (ozone). Any values missing in the original dataset were imputed via predictive mean matching \citep{rubin1986statistical}. We also artificially generated two static variables from the standard half-normal distribution.

The full Chicago data has 5,114 observations. In order to test the models under different sample sizes, we used two sets of simulations: one that contains all 5,114 observations, and another that only contains the first 250.

\begin{figure}
	\centering
	\includegraphics[scale=0.4]{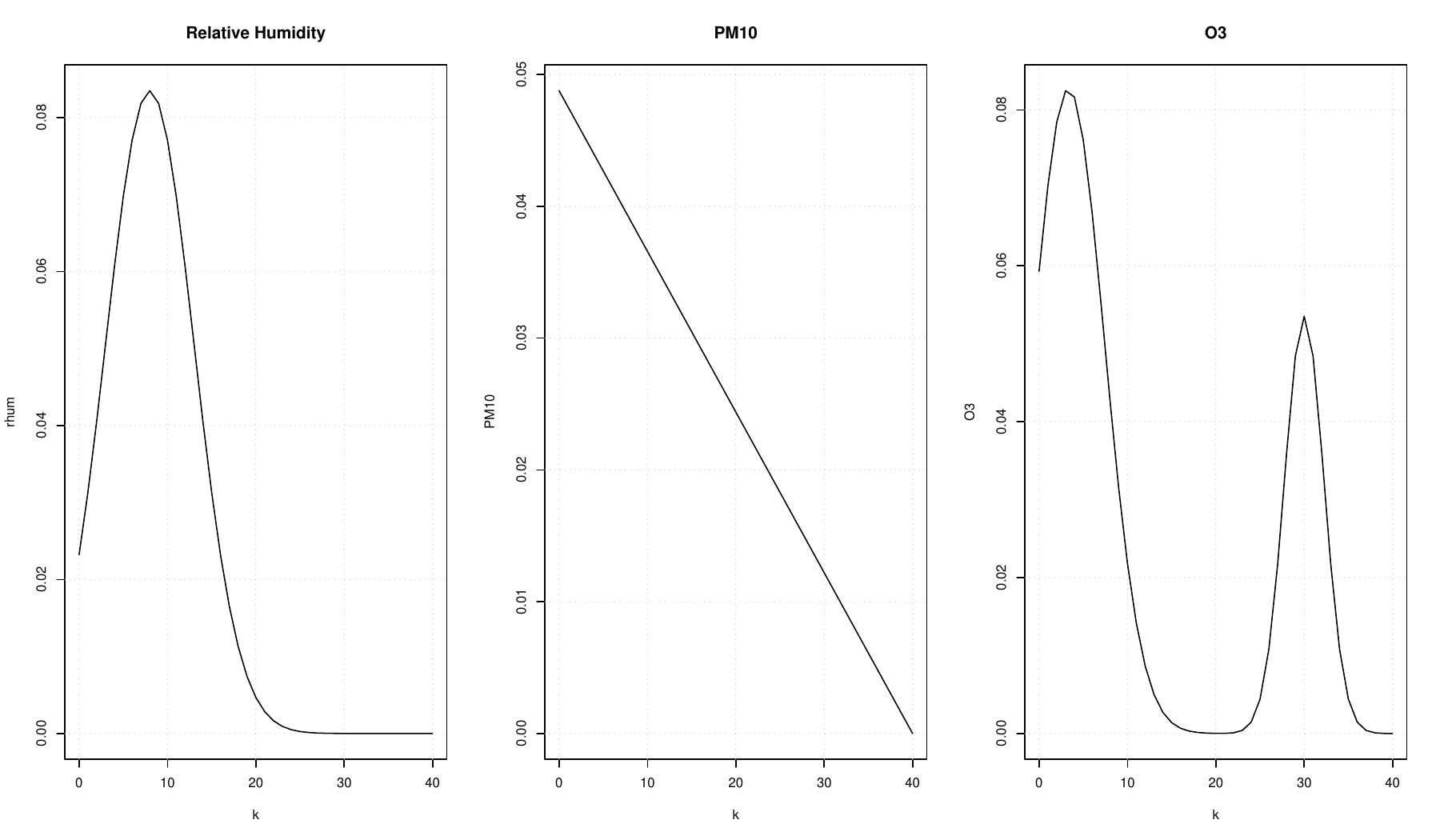}
	\caption{\label{fig:sim_lr}The scaled lag-response curves used in the simulations for each of the dynamic predictors.}
\end{figure}

In order to simulate a response variable, we first need to specify the lag-response for the dynamic predictors. In order to have a variety of effects to test, we gave each predictor a unique kind of lag-response. For rhum and O3, we set the lag-response using the DLF method explained in Section~\ref{sec:method}. More specifically, we used a 2-degree exponential Almon function \citep{almon1965distributed}:

\begin{equation} \label{eq:nealmon}
	w(k, \boldtheta) \propto \exp\left( \theta_1 k + \theta_2 k^2 \right)
\end{equation}

\noindent where $w(\cdot, \cdot)$ is scaled across all lags $k$ within each time window such that $\sum_{k=0}^\tau w(k, \boldtheta) = 1$. For rhum, we used $\boldtheta = \{0.32, -0.02\}$. For O3 we wanted a bimodal effect, so to accomplish this we took two Almon curves, with the first curve $w_1$ having parameters $\boldtheta = \{0.2, -0.03\}$ and the second $w_2$ having parameters $\boldtheta = \{6, -0.1\}$. The weights for O3 were then taken to be $w = 0.7w_1 + 0.3w_2$. As for PM10, we assigned it a linearly decreasing lag-response. Every dynamic predictor used 40 lags. The shape of these lag-responses are shown in Figure~\ref{fig:sim_lr}.

The overall effect sizes were set as follows: the two static variables have an overall effect of $-0.5$ and $0.01$. Rhum, PM10 and O3 were assigned overall effects sizes of $0.5$, $0.01$, and $-0.5$ respectively. This gives us a mix of strong and weak effects to test the variable selection scheme.

Datasets were created with both a count response and a binary response. The count response simulations used the negative binomial distribution with stopping parameter equal to $50$, and intercept equal to $-1$. The binary response variable was simulated by drawing random values from the ALD with the skew parameter equal to $0.9$ and setting the intercept of the regression model to 0.

In summation, there are four distinct types of simulations; Large/small datasets with count/binary responses. To asses Monte Carlo consistency, four of each type of dataset was created, and for each of these datasets we fit the appropriate model 30 times.

When fitting the models to the simulation data, we ran the MCMC chains for 50,000 iterations, discarding the first half as burn-in. Every $10^\text{th}$ value was retained thereafter. For modelling the lag-response, we did not use the same DLF method that we used to simulate the data; instead, we used a cubic B-spline with three evenly spaced knots (i.e., knots at the $25^\text{th}$, $50^\text{th}$ and $75^\text{th}$ percentiles) along the lags. The basis matrices therefore add seven columns to the design matrix for each of the three dynamic variables.

\begin{figure}
	\centering
	\begin{subfigure}[b]{1\textwidth}
		\centering
		\includegraphics[width=\textwidth]{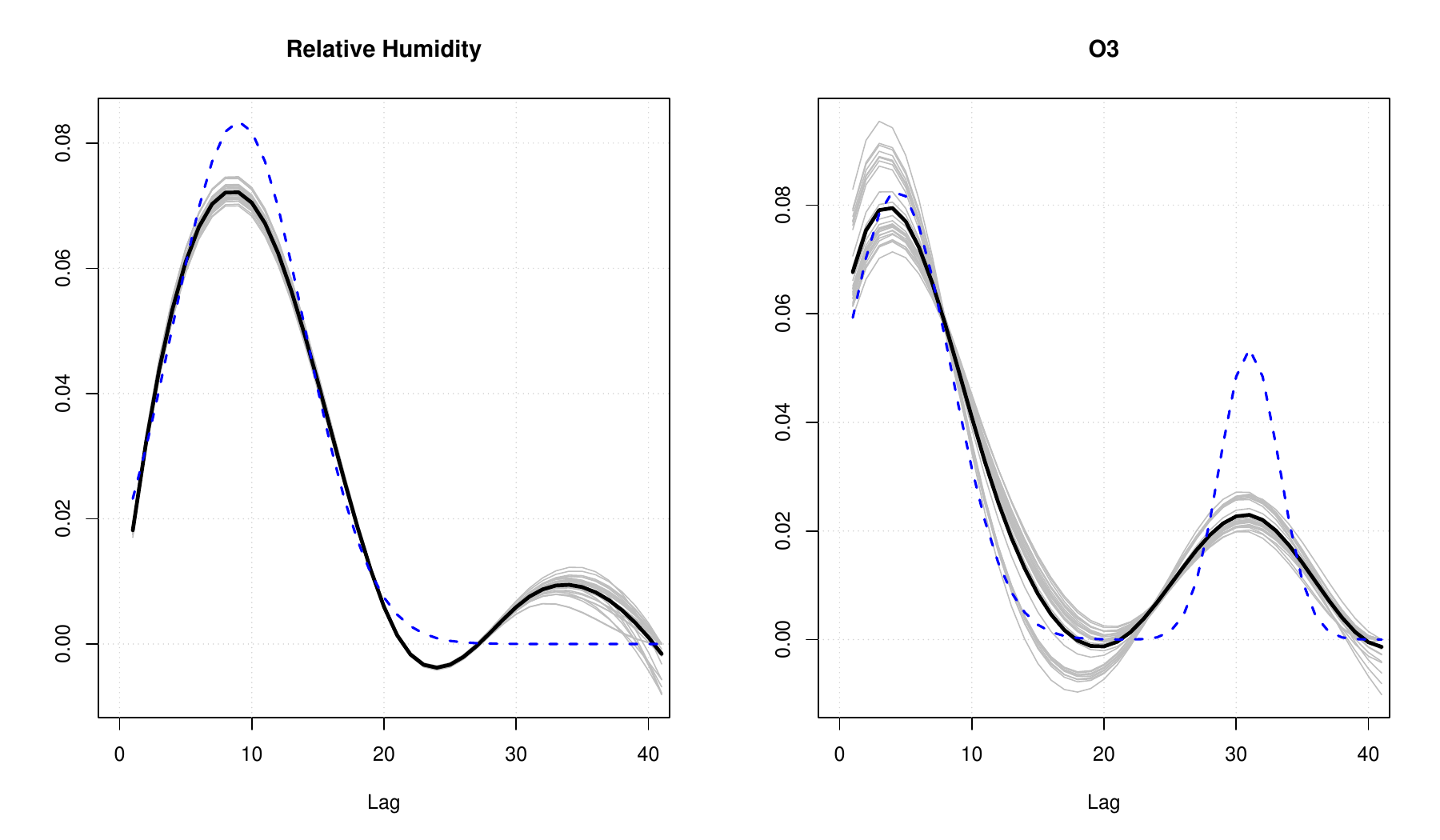}
		\caption{Count data simulation results.}
	\end{subfigure}
	\hfill
	\begin{subfigure}[b]{1\textwidth}
		\centering
		\includegraphics[width=\textwidth]{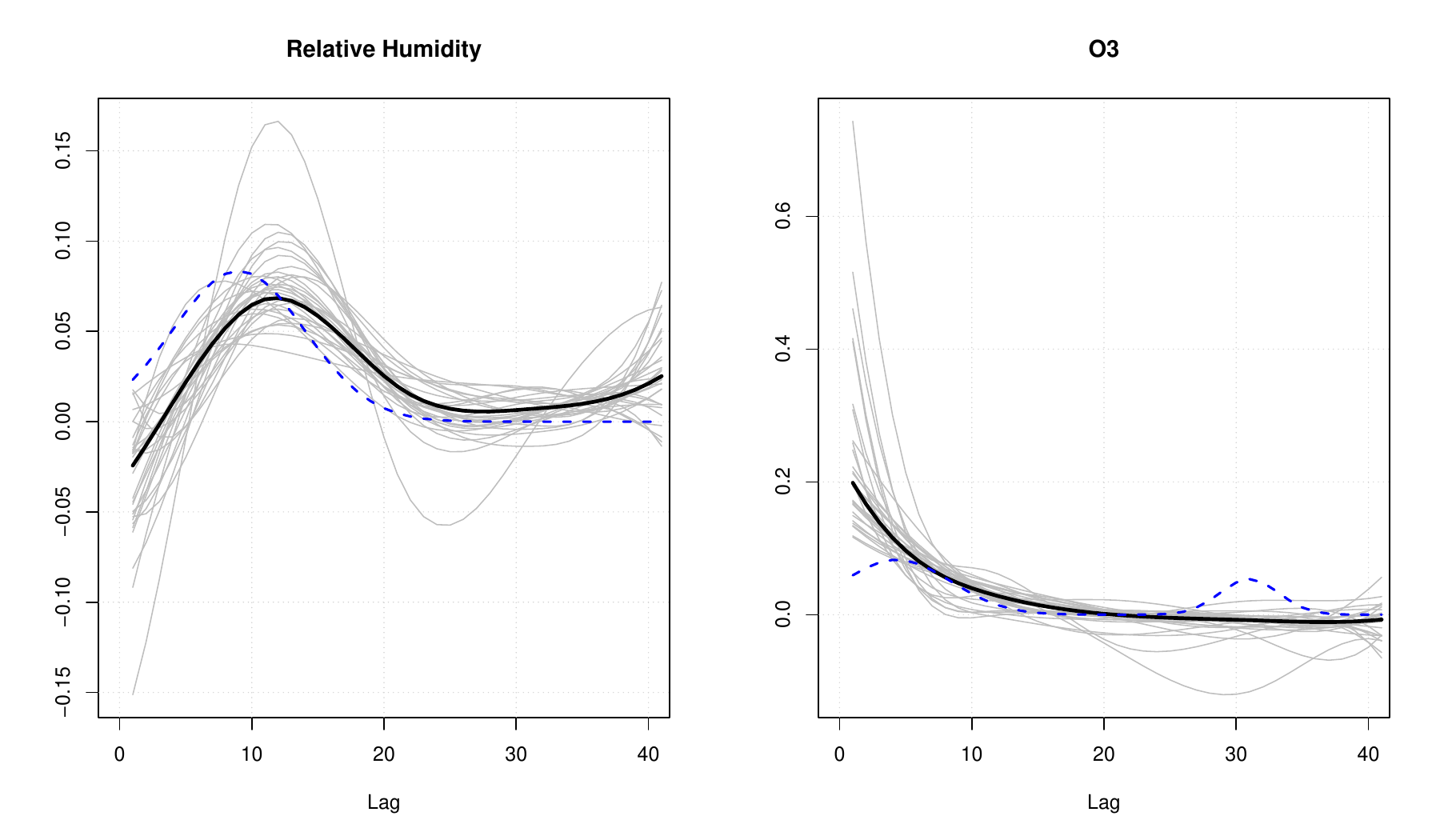}
		\caption{Binary data simulation results.}
	\end{subfigure}
	\caption{Estimated lag-responses for simulated data compared to their true values. Grey lines represent each of the 30 MCMC runs, the thick black line is the average across all runs, and the blue dashed lines are the correct values. The correct values are the same for both groups, they only appear different here due to the differing scales of the y-axes. PM10 is omitted here as its small effect size means that its associated lag-response is ill-defined, though it is included in the supplementary results.}
	\label{fig:sim_DLF}
\end{figure}

For brevity, the only simulation results we will discuss here are one (out of four) of the large simulations for both count and binary simulations; see Figures~\ref{fig:sim_DLF}--\ref{fig:sim_if}. Relevant visualisations for \emph{all} simulation tests we ran are compiled in the supplementary material.

\begin{figure}
	\centering
	\begin{subfigure}[b]{1\textwidth}
		\centering
		\includegraphics[width=\textwidth]{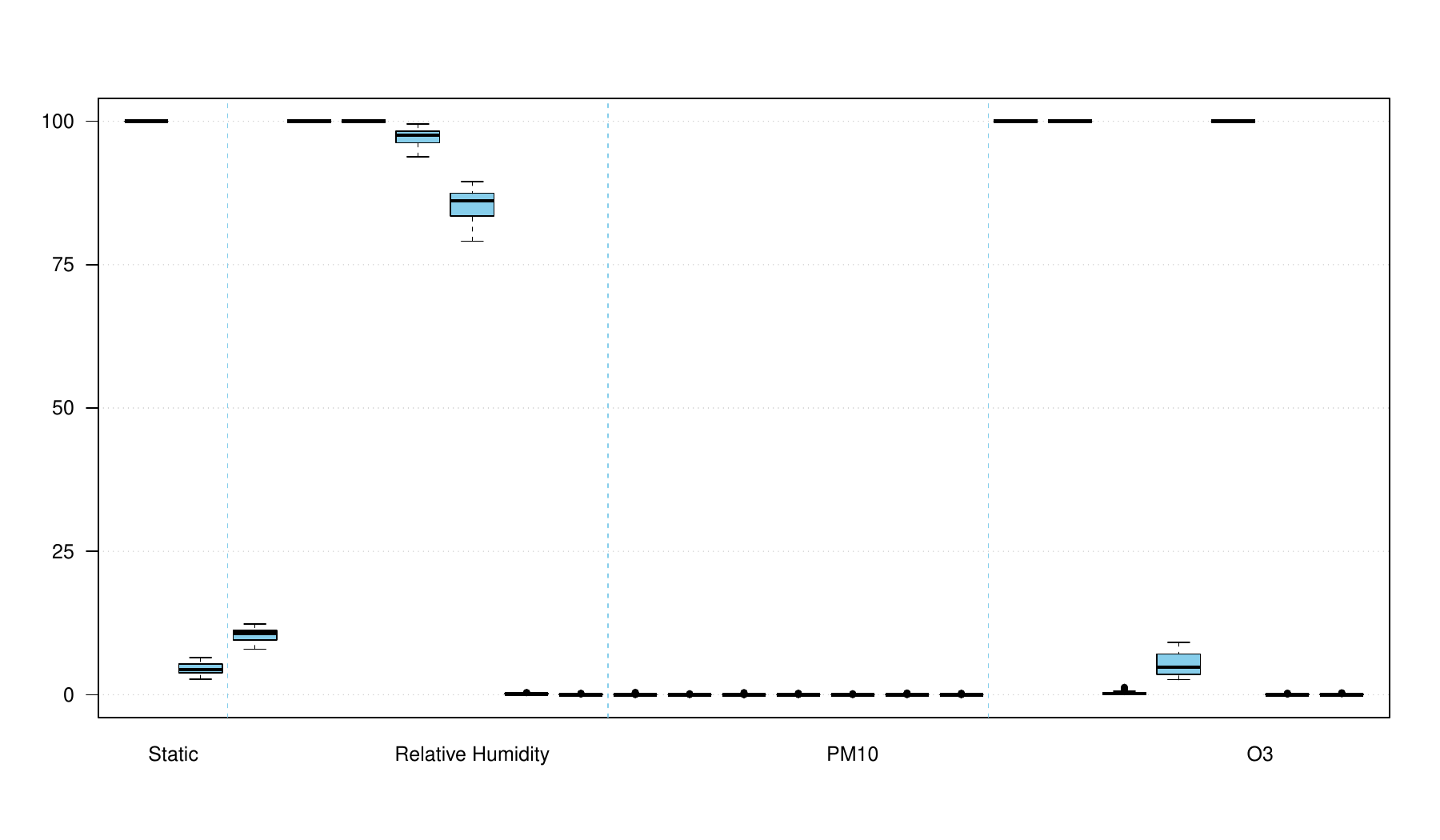}
		\caption{Count data simulation results.}
	\end{subfigure}
	\hfill
	\begin{subfigure}[b]{1\textwidth}
		\centering
		\includegraphics[width=\textwidth]{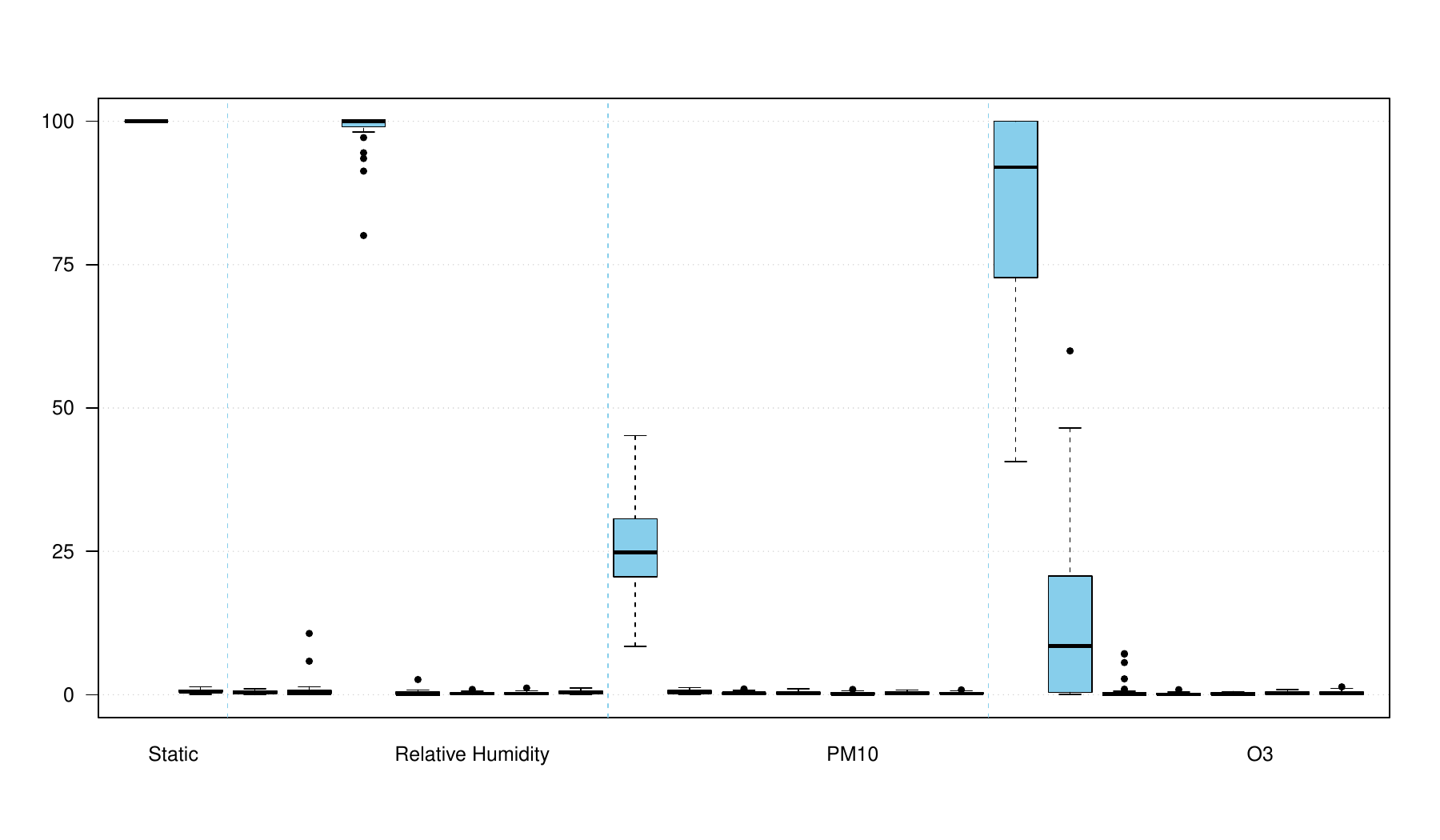}
		\caption{Binary data simulation results.}
	\end{subfigure}
	\caption{\label{fig:sim_gamma}Covariate inclusion posteriors for simulated data across all 30 MCMC runs (intercept omitted).}
\end{figure}

\begin{figure}
	\centering
	\begin{subfigure}[b]{1\textwidth}
		\centering
		\includegraphics[width=\textwidth]{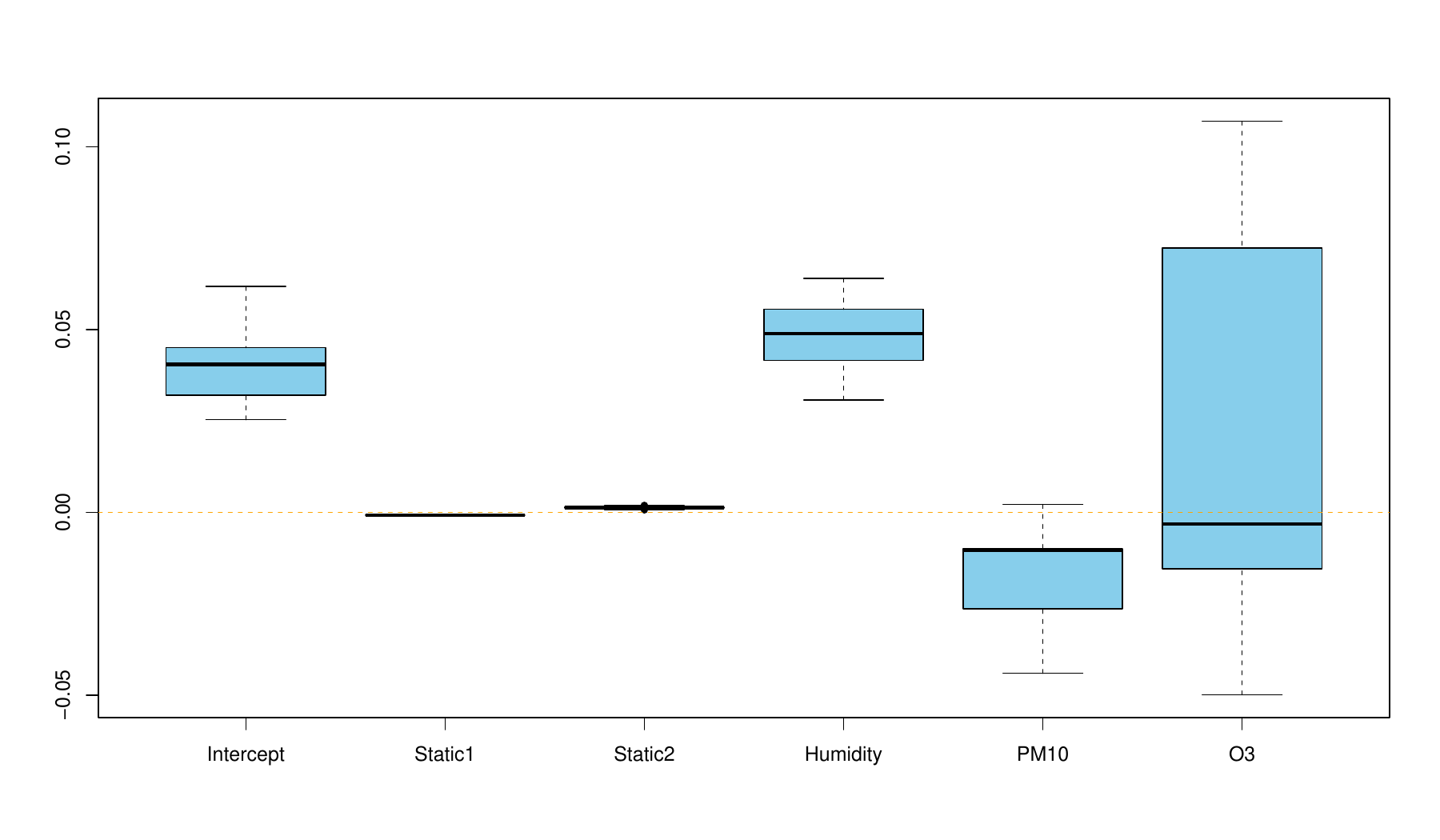}
		\caption{Count data simulation results.}
	\end{subfigure}
	\hfill
	\begin{subfigure}[b]{1\textwidth}
		\centering
		\includegraphics[width=\textwidth]{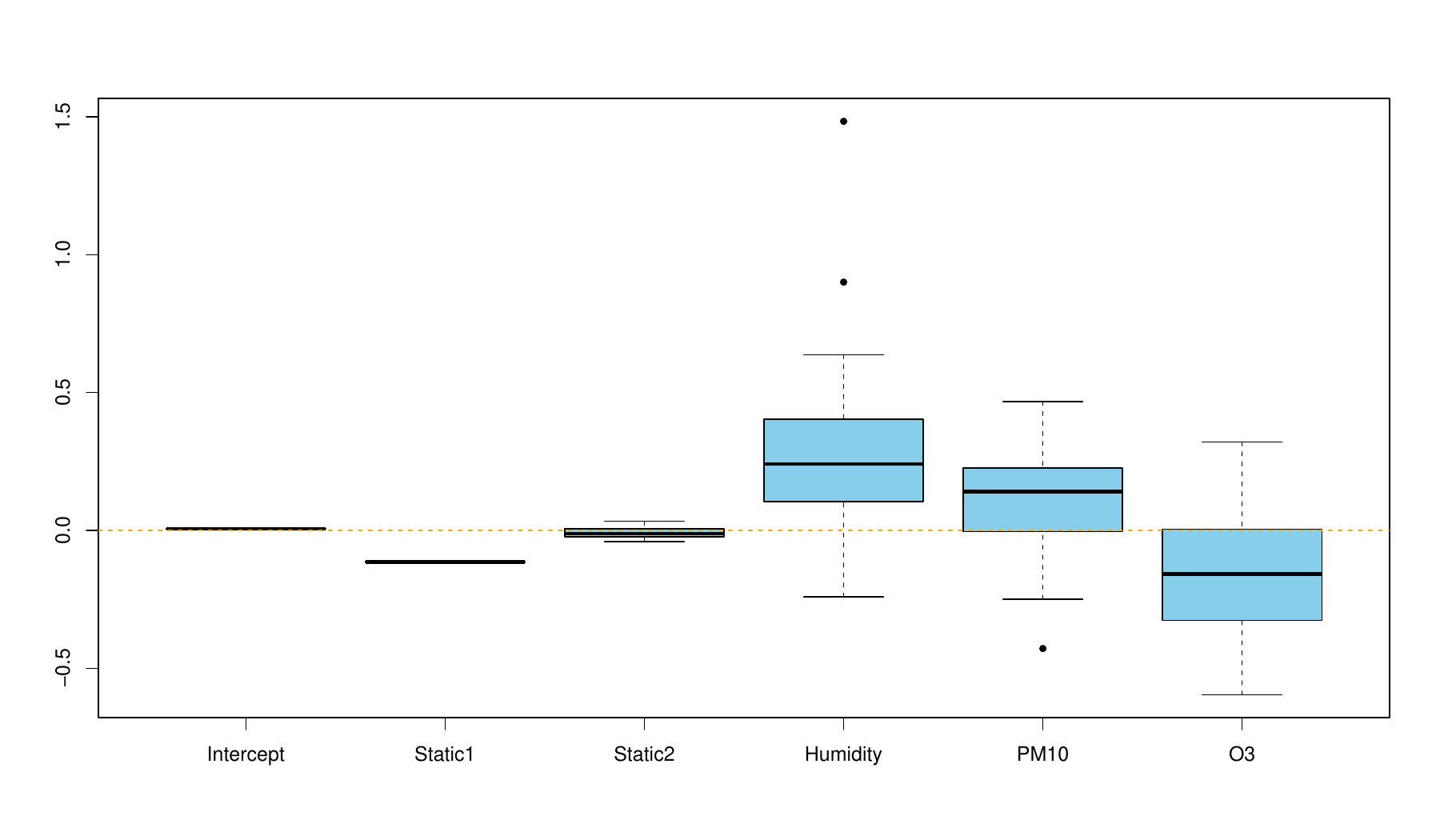}
		\caption{Binary data simulation results.}
	\end{subfigure}
	\caption{\label{fig:sim_beta}The difference between the mean estimated effect sizes and their true values for simulated data across all 30 MCMC runs. Note that the y-axes of the above graphs are on different scales.}
\end{figure}

\begin{figure}
		\centering
		\includegraphics[scale=0.4]{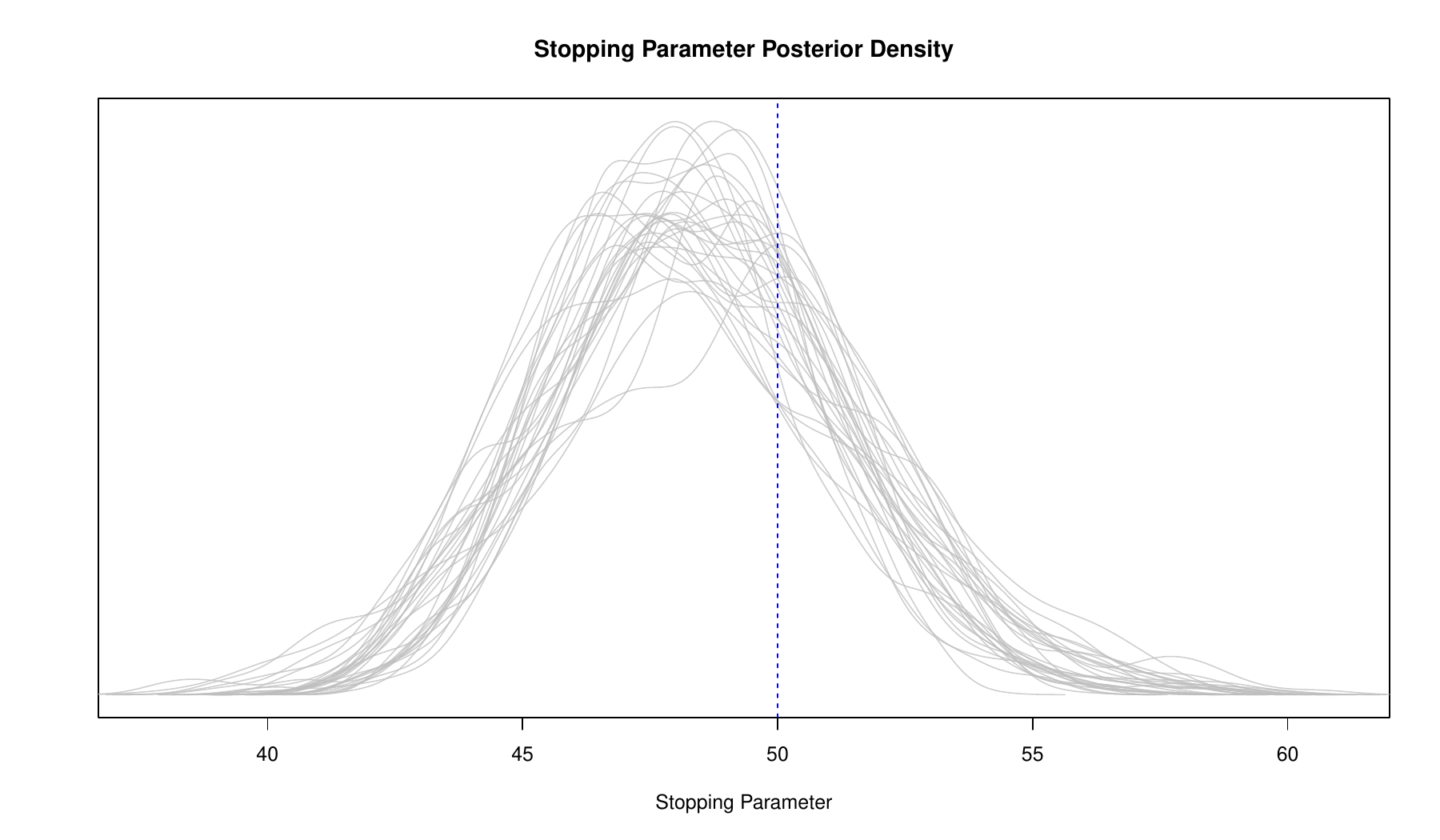}
	\caption{\label{fig:sim_xi}Posterior of the negative binomial stopping parameter for simulated data. Each grey line represents a posterior from one of the thirty MCMC runs. The blue dashed line denotes the true value.}
\end{figure}

\begin{figure}
	\centering
	\begin{subfigure}[b]{1\textwidth}
		\centering
		\includegraphics[width=\textwidth]{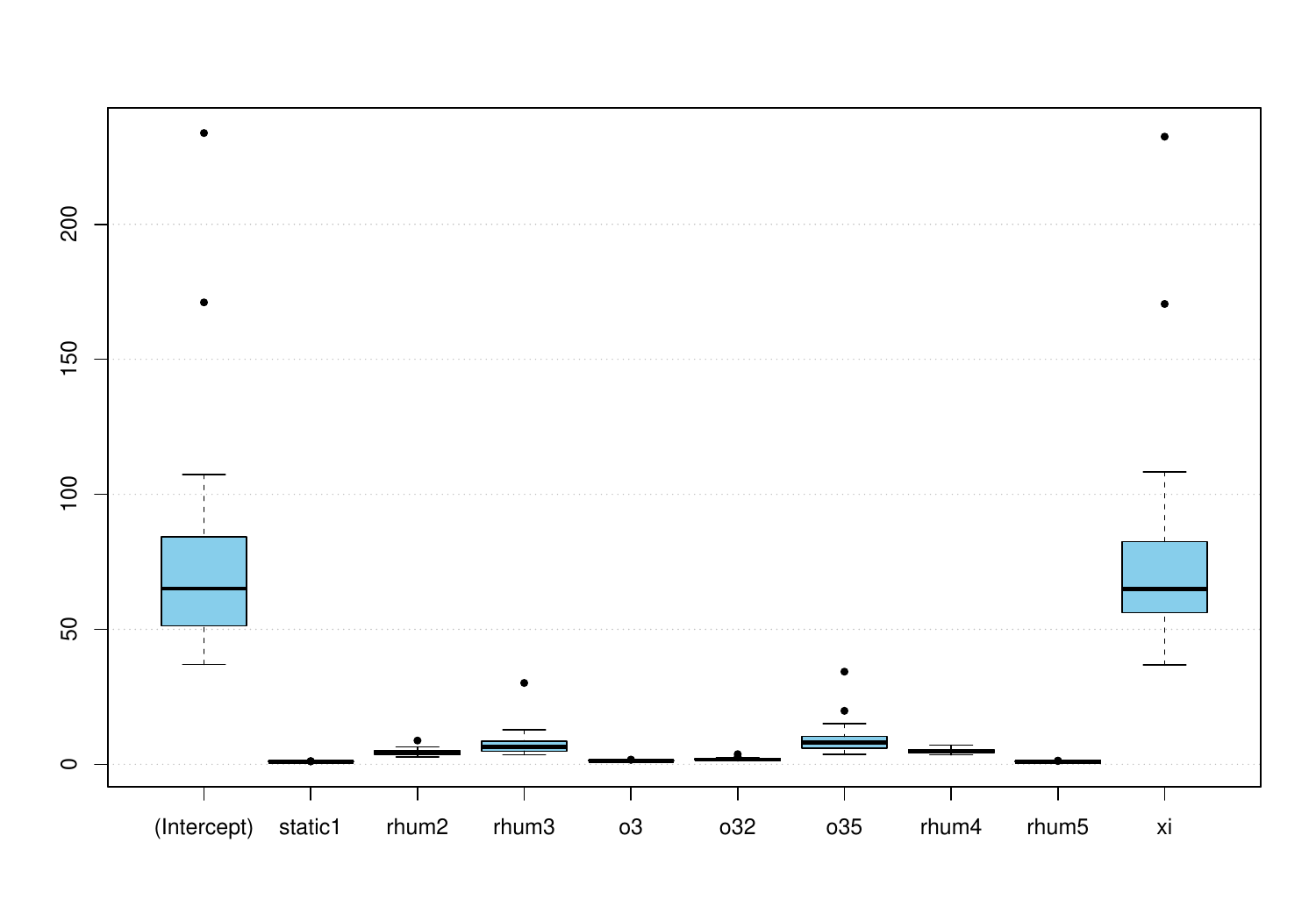}
		\caption{Count data simulation results.}
	\end{subfigure}
	\hfill
	\begin{subfigure}[b]{1\textwidth}
		\centering
		\includegraphics[width=\textwidth]{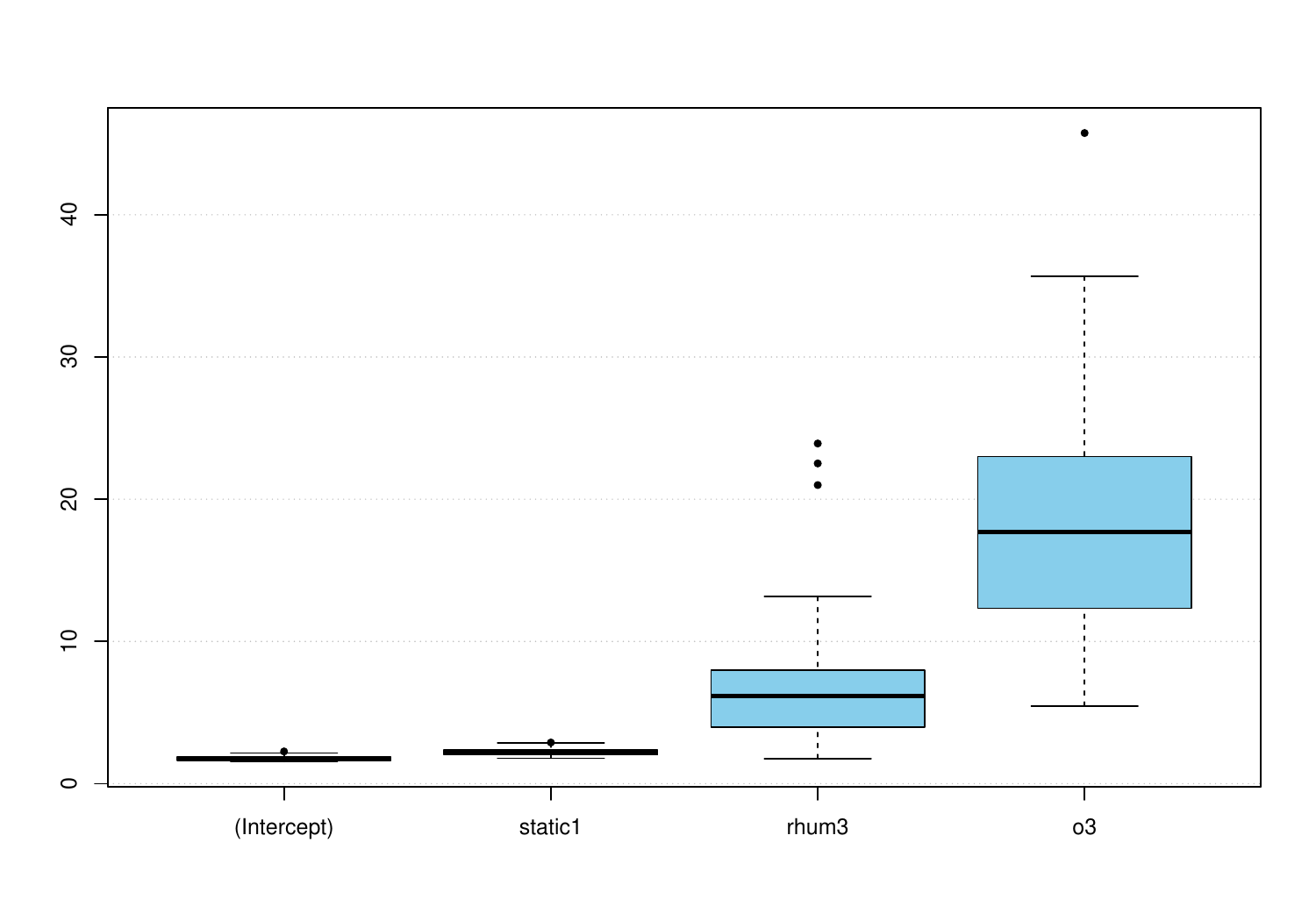}
		\caption{Binary data simulation results.}
	\end{subfigure}
	\caption{\label{fig:sim_if}The Inefficiency Factor across all 30 runs. We only include predictors here that had more than 50\% inclusion across all 30 runs. As with the graphs above, note the differences in y-axis scale between count and binary simulation results.}
\end{figure}

From Figure~\ref{fig:sim_DLF} we see that the MCMC routine was able to accurately infer the lag-response relationship for relative humidity and ozone for the count-response data. Binary simulations were somewhat less successful; the lag-response relationship with relative humidity, while overall accurate, is less consistent, and the algorithm failed to pick up the smaller mode for the ozone lag-response relationship. PM10 was omitted as the overall effect is so small that the lag-response is ill-defined and thus expected to fail. However the attempt at the estimation is shown in the supplementary materials. 

Figure~\ref{fig:sim_gamma} shows the variable inclusion parameters. The weaker effects are largely discarded from the model with a high degree of certainty. More interestingly, we see that $\boldgamma$ are discarding individual B-spline components even from the non-zero effects. This behaviour could be potentially leveraged to inform about the parameters of the b-spline construction (knots, component polynomial degree, etc). The binary response simulations, while performing well overall, show a lack of sampling consistency, especially with the ozone variable.

The difference between estimated and true effect sizes are shown in Figure~\ref{fig:sim_beta}. We see that the effects were mostly accurate, though some effects were overestimated, in particular the intercept and humidity slope in the negative binomial simulations. The other simulations we performed, shown in the supplementary materials, show further varying degrees of over/under estimation for some dynamic effects for both count and binary response simulations. By contrast, the static effects (apart from the intercept) are almost always very accurate with a high degree of Monte Carlo consistency. 

Another thing worth noting here are the differences in the count and binary simulations - as before, imbalanced binary response is harder to infer from (note the difference in y-axes scales in Figure~\ref{fig:sim_beta}). 

Figure \ref{fig:sim_xi} displays the density plot of the MCMC posterior samples of the negative binomial stopping parameter. The Monte Carlo posterior mode tends to underestimate, but the true value still sits comfortably within high posterior density regions. 

To assess MCMC convergence, Figure~\ref{fig:sim_if} displays the Inefficiency Factor (IF) \citep{chib2011introduction}. We only display predictors that had more than 50\% inclusion across all 30 runs. Binary regression performs decently by this metric, with only one predictor typically exceeding an IF of 10. However, the count simulations reveal an excessive IF for the intercept and stopping parameter $\xi$, indicating poor convergence for those parameters, despite their relative accuracy in estimation. This is caused by a large degree of mutual colinearity in their posterior samples; the Pearson correlation between the intercept and $\xi$ samples are -0.997 for the first run. This is likely due to the fact that the mean of the negative binomial and the stopping parameter both directly contribute to the variance; thus as one changes, the other must shift accordingly to compensate for the observed variance.

Overall, we would say the negative binomial simulations performed better at parameter estimation, and with typically better consistency across runs, but the Markov Chain suffers diagnostic issues. The quantile binary model was still reliable at variable selection and didn't appear to suffer poor convergence diagnostics.

\section{Chicago Air Quality Data} \label{sec:data}
We now fit DLM models to the Chicago air quality dataset that was used to partially simulate response data in the previous section. This data was originally part of the National Morbidity, Mortality and Air Pollution Study database, but to our knowledge the full dataset is no longer available. The data we use here is available through the \emph{dlnm} package \citep{gasparrini2011distributed}. In addition to the relative humidity, PM10 levels, and ozone levels that we spoke of in the simulation section, the full dataset has two more environmental variables: mean daily temperature and dew point temperature, both of which are heavily correlated (Pearson $\approx$ 0.95). It also records the day of the week and the month of the year the air quality metrics were recorded. Month of the year is obviously useful for coding seasonality, and day of week is also known to have an impact on daily reported numbers due to the nature of healthcare surveillance \citep{Buckingham2017correcting}. Day of the week will be dummy-encoded and included in the model (Sunday excluded). The months were aggregated into their respective seasons, which was dummy encoded and included in the model (Autumn excluded).

There are three different measures of daily death rates in the data: the number of cardiovascular-related deaths, the number of respiratory related deaths, and the overall number of non-accidental deaths. The daily overall death rate was chosen as the response; the others were discarded. As in the simulation experiments, missing values are imputed using predictive mean matching. 

The remaining air quality metrics are included as dynamic variables. These are: mean temperature, dew point temperature, mean relative humidity, particulate matter (PM10), and ozone (O3). In each case we assume a lag length of 40 days, and we use a cubic B-spline to form the lag-response, as we did previously in Section~\ref{sec:sim_study}. As before, we used 3 knots that were spread evenly across the lags, at the $25^\text{th}$, $50^\text{th}$, and $75^\text{th}$ percentiles.

We fit a negative binomial DLM to the dataset as described. We also fit a binary quantile regression DLM to the dataset by dichotimising the response variable; any day that had in excess of 135 deaths (roughly 10\% of the data) was assigned a value of 1, and 0 otherwise.

For the MCMC fits shown, we adopt independent Gaussian distributions centered at 0 with variance 100 as priors for the regression coefficients. We use a Bernoulli distribution with probability parameter equal to 0.5 as a prior for each of the variable inclusion parameters. A Gamma distribution with shape and rate equal to 2 and 1/50 respectively was used for the stopping parameter prior. We run the MCMC algorithm 4 times each with their own starting values. One run starts with the intercept only model, another starts at the full inclusion model, and the other two use random starts; each non-intercept predictor is included with probability 0.5. They are all ran for $10^5$ iterations, half of which is discarded, and every $10^\text{th}$ value is retained thereafter.

\begin{figure}
	\centering
	\begin{subfigure}[b]{1\textwidth}
		\centering
		\includegraphics[scale=0.4]{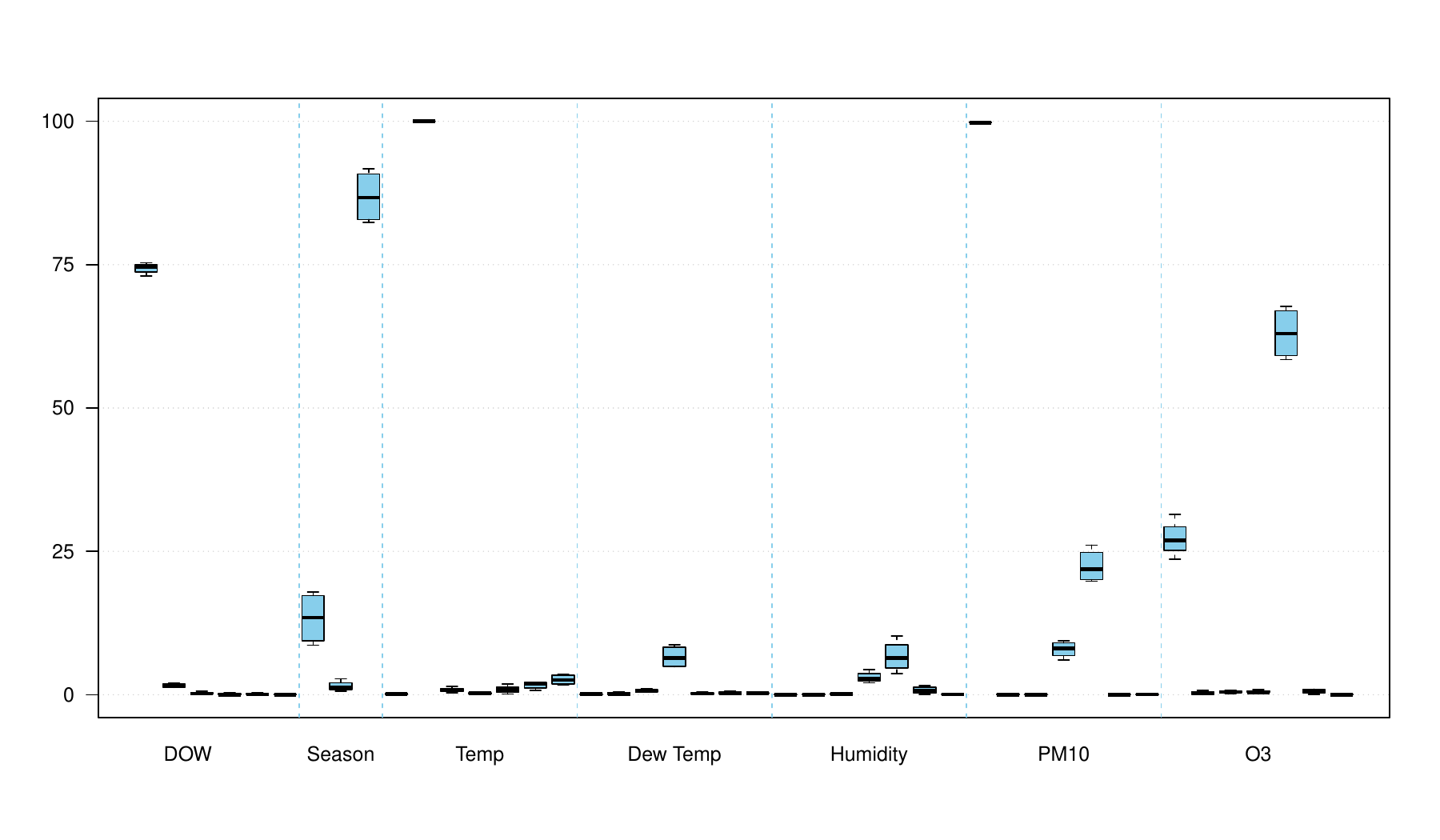}
		\caption{Covariate inclusion posterior across all 4 runs.}
	\end{subfigure}
	\hfill
	\begin{subfigure}[b]{1\textwidth}
		\centering
		\includegraphics[width=\textwidth]{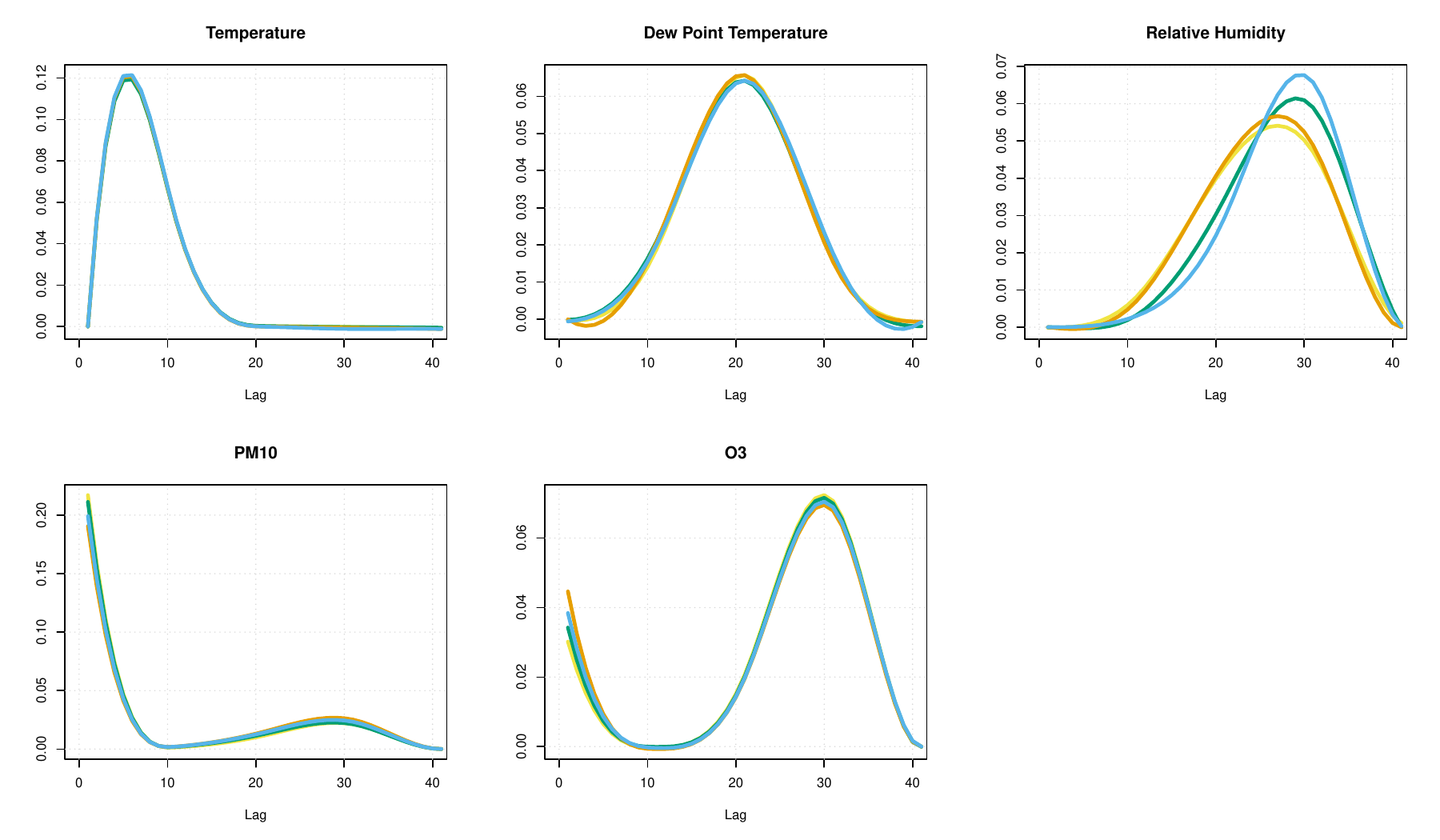}
		\caption{Monte Carlo means of the response-lag relationship for all 4 runs.}
	\end{subfigure}
	\caption{Results from the negative binomial fit of the Chicago air quality data.}
	\label{fig:dat_res_nb}
\end{figure}

\begin{figure}
	\centering
	\begin{subfigure}[b]{1\textwidth}
		\centering
		\includegraphics[scale=0.4]{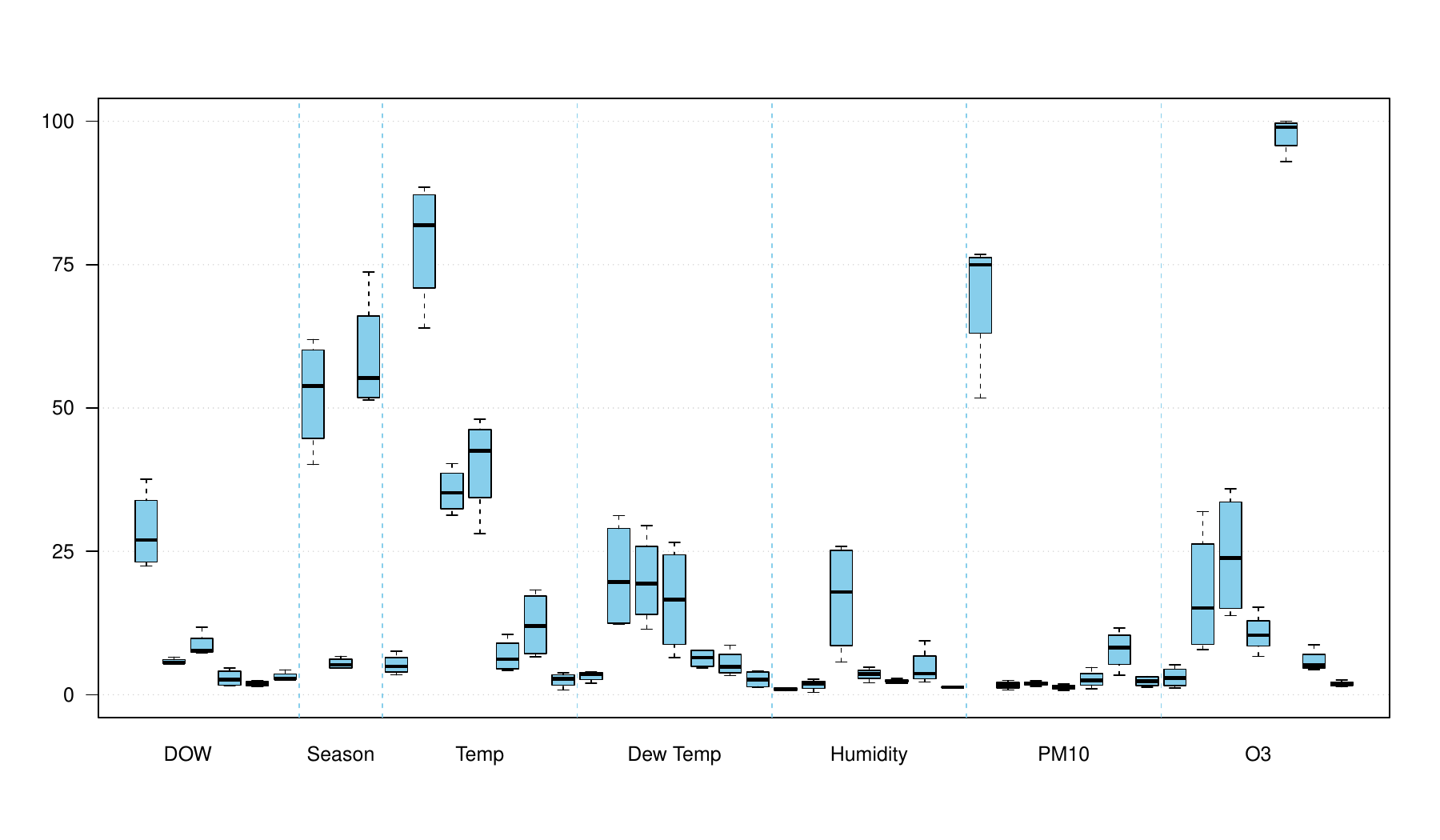}
		\caption{Covariate inclusion posterior across all 4 runs.}
	\end{subfigure}
	\hfill
	\begin{subfigure}[b]{1\textwidth}
		\centering
		\includegraphics[width=\textwidth]{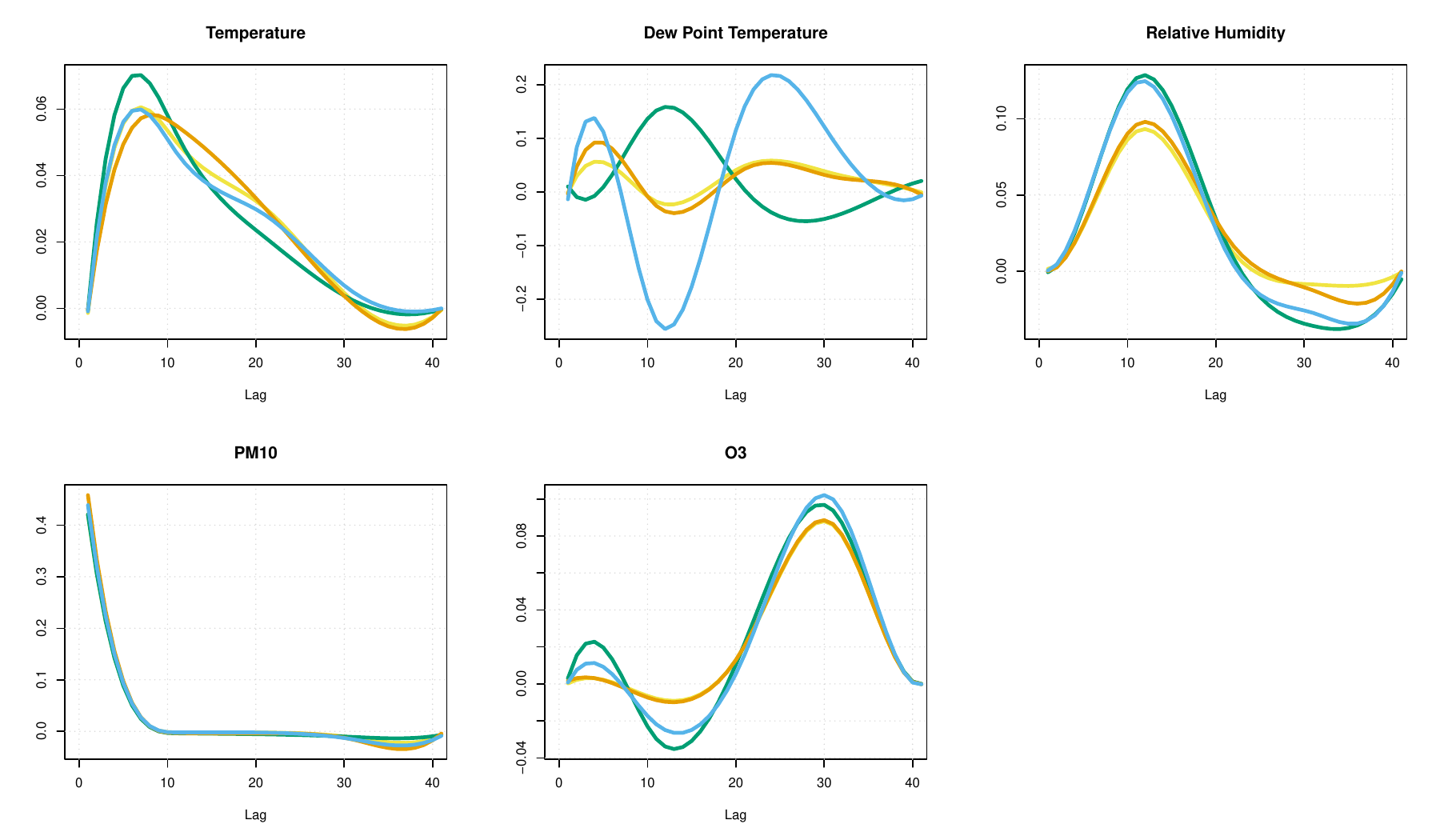}
		\caption{Monte Carlo means of the lag-response relationship for all 4 runs.}
	\end{subfigure}
	\caption{Results from the binary quantile regression fit of the Chicago air quality data.}
	\label{fig:dat_res_qr}
\end{figure}

The posterior of the covariate inclusion parameters and the means of the estimated lag-response relationships are given in Figures~\ref{fig:dat_res_nb} and \ref{fig:dat_res_qr}. The negative binomial DLM shows very consistent results for both the inclusion indicators and the lag-responses, even with the large degree of multicolinearity between temperature and dew-point temperature. The distribution of variable inclusions (top panel of Figure~\ref{fig:dat_res_nb}) affirms the day-of-week effect and a seasonal effect on the rates of the daily death rates. Of the dynamic predictors, temperature, PM10, and ozone appear to have influence according to our findings. The bottom panel, showing the inferred lag-response of the dynamic variables, indicates that the effect of temperature is delayed by roughly a week, PM10 has a very sharp immediate effect, and that the effect of ozone is bimodal; an immediate effect, as well as a month-delayed effect.

There was more uncertainty within the quantile regression results (Figure~\ref{fig:dat_res_qr}). The day-of-week effect appears suppressed here, but the seasonal effect still appears to be present. In terms of the dynamic predictors, it appears that there is a lot of agreement with the negative binomial model above, albeit with more uncertainty; temperature, PM10 and ozone have a large degree of inclusion. In the bottom panel, we see that temperature has a more immediate effect than it did for the count model, but the fatter tail implies that the negative effects of temperature are slightly more persistent within the 40-day lag window. Ozone's immediate effect is diminished compared to the count model, but it's month-delayed effect is far more pronounced. PM10 appears to have a strong immediate effect, as before.

\section{Conclusion} \label{sec:conclusion}
Latent variables are frequently leveraged when fitting Bayesian GLMs (especially binary response models) as they result in fully tractable conditional distributions and require no further tuning beyond setting priors and starting values for MCMC. Here we focused on the implementation of quantile binary regression and negative binomial regression, but any similar latent variable approach can be incorporated just as easily. For example, \cite{rahman2016bayesian} and \cite{maheshwari2023bqror} explain how binary quantile regression can be readily expanded to fit a quantile ordinal regression model. A known downside of these approaches, beyond the fact that some carry a large computational cost, is that in certain situations (such as when binary data is heavily imbalanced) there can be severe autocorrelation in the MCMC posterior samples due to the close relationship between the parameters and the latent variables, and may be outperformed by a Metropolis-Hastings approach \citep{johndrow2018mcmc}. \cite{zens2023ultimate} proposed using parameter expansion \citep{liu1999parameter} to improve the efficiency of fitting a logistic regression model via P\'olya-Gamma latent variables. \cite{duan2018scaling} suggest improvements to the efficiency by scaling the conditional posterior variance; these methods and others can be incorporated to the DLM framework used here if desired.


All the code used for this study can be found within the following link: \url{https://github.com/DanDempsey/Latent-Variable-DLM-Project/tree/main/Code}. We have also developed an \texttt{R} package that implements the methods discussed in this paper. In can be installed through the following command in \texttt{R}: \\\texttt{devtools::install\_github("DanDempsey/DiscreteDLM")}.

\section*{Acknowledgments}
Daniel Dempsey's work was funded by an Irish Research Council Enterprise Partnership Scheme Award EPSPG/2017/304. Jason Wyse's work was partially supported through a Science Foundation Ireland Frontiers for the Future Award 21/FFP$-$P/10123 and the European Union’s Horizon 2020 research and innovation programme under the Marie Skłodowska-Curie grant agreement No. 813545.

\subsection*{Disclosure Statement}
The authors have no competing interests to declare.

\bibliography{Bayes_Quantile_DLM_bib}

\begin{thebibliography}{}

\bibitem[Albert and Chib, 1993]{albert1993}
Albert, J.~H. and Chib, S. (1993).
\newblock Bayesian {{Analysis}} of {{Binary}} and {{Polychotomous Response
  Data}}.
\newblock {\em Journal of the American Statistical Association},
  88(422):669--679.

\bibitem[Almon, 1965]{almon1965distributed}
Almon, S. (1965).
\newblock The distributed lag between capital appropriations and expenditures.
\newblock {\em Econometrica: Journal of the Econometric Society}, pages
  178--196.

\bibitem[Antonelli et~al., 2021]{antonelli2021multiple}
Antonelli, J., Wilson, A., and Coull, B. (2021).
\newblock Multiple exposure distributed lag models with variable selection.

\bibitem[Benoit and den Poel, 2017]{benoit2017}
Benoit, D.~F. and den Poel, D.~V. (2017).
\newblock {{bayesQR}}: {{A Bayesian Approach}} to {{Quantile Regression}}.
\newblock {\em Journal of Statistical Software}, 76(1):1--32.

\bibitem[Buckingham-Jeffery et~al., 2017]{Buckingham2017correcting}
Buckingham-Jeffery, E., Morbey, R., House, T., Elliot, A.~J., Harcourt, S., and
  Smith, G.~E. (2017).
\newblock Correcting for day of the week and public holiday effects: improving
  a national daily syndromic surveillance service for detecting public health
  threats.
\newblock {\em BMC Public Health}, 17(1):1--9.

\bibitem[Carvalho et~al., 2010]{carvalho2010horshoe}
Carvalho, C.~M., Polson, N.~G., and Scott, J.~G. (2010).
\newblock The horseshoe estimator for sparse signals.
\newblock {\em Biometrika}, 97(2):465--480.

\bibitem[Chib, 2011]{chib2011introduction}
Chib, S. (2011).
\newblock Introduction to {Simulation} and {MCMC} {Methods}.
\newblock In {\em The Oxford handbook of Bayesian econometrics}, pages
  183--218. Oxford University Press, Oxford.

\bibitem[Czado and Santner, 1992]{czado1992effect}
Czado, C. and Santner, T.~J. (1992).
\newblock The effect of link misspecification on binary regression inference.
\newblock {\em Journal of statistical planning and inference}, 33(2):213--231.

\bibitem[D'Angelo and Canale, 2022]{dangelo2022efficient}
D'Angelo, L. and Canale, A. (2022).
\newblock Efficient posterior sampling for {Bayesian} {Poisson} regression.
\newblock {\em arXiv preprint arXiv:2109.09520}.

\bibitem[DiMatteo et~al., 2001]{dimatteo2001bayesian}
DiMatteo, I., Genovese, C.~R., and Kass, R.~E. (2001).
\newblock Bayesian curve-fitting with free-knot splines.
\newblock {\em Biometrika}, 88(4):1055--1071.

\bibitem[Duan et~al., 2018]{duan2018scaling}
Duan, L.~L., Johndrow, J.~E., and Dunson, D.~B. (2018).
\newblock Scaling up data augmentation mcmc via calibration.
\newblock {\em The Journal of Machine Learning Research}, 19(1):2575--2608.

\bibitem[Eilers and Marx, 2010]{eilers2010splines}
Eilers, P.~H. and Marx, B.~D. (2010).
\newblock Splines, knots, and penalties.
\newblock {\em Wiley Interdisciplinary Reviews: Computational Statistics},
  2(6):637--653.

\bibitem[Foroni et~al., 2015]{foroni2015unrestricted}
Foroni, C., Marcellino, M., and Schumacher, C. (2015).
\newblock Unrestricted mixed data sampling ({MIDAS}): {MIDAS} regressions with
  unrestricted lag polynomials.
\newblock {\em Journal of the Royal Statistical Society: Series A (Statistics
  in Society)}, 178(1):57--82.

\bibitem[Fr{\"u}hwirth-Schnatter and Fr{\"u}hwirth, 2010]{fruhwirth2010data}
Fr{\"u}hwirth-Schnatter, S. and Fr{\"u}hwirth, R. (2010).
\newblock Data augmentation and {MCMC} for binary and multinomial logit models.
\newblock {\em Statistical modelling and regression structures: Festschrift in
  honour of Ludwig Fahrmeir}, pages 111--132.

\bibitem[Gasparrini, 2011]{gasparrini2011distributed}
Gasparrini, A. (2011).
\newblock Distributed lag linear and non-linear models in {R}: the package
  {dlnm}.
\newblock {\em Journal of Statistical Software}, 43(8):1--20.

\bibitem[Gasparrini, 2013]{gasparrini2013modeling}
Gasparrini, A. (2013).
\newblock Modeling exposure--lag--response associations with distributed lag
  non-linear models.
\newblock {\em Statistics in medicine}, 33(5):881--899.

\bibitem[Gasparrini et~al., 2010]{gasparrini2010distributed}
Gasparrini, A., Armstrong, B., and Kenward, M.~G. (2010).
\newblock Distributed lag non-linear models.
\newblock {\em Statistics in medicine}, 29(21):2224--2234.

\bibitem[Ghysels et~al., 2016]{ghysels2016}
Ghysels, E., Kvedaras, V., and Zemlys, V. (2016).
\newblock Mixed {{Frequency Data Sampling Regression Models}}: {{The R
  Package}} midasr.
\newblock {\em Journal of Statistical Software}, 72(1):1--35.

\bibitem[Ghysels et~al., 2004]{ghysels2004midas}
Ghysels, E., Santa-Clara, P., and Valkanov, R. (2004).
\newblock The {MIDAS} touch: Mixed data sampling regression models.
\newblock {\em UCLA: Finance}.

\bibitem[Ghysels et~al., 2007]{ghysels2007midas}
Ghysels, E., Sinko, A., and Valkanov, R. (2007).
\newblock {MIDAS} regressions: {Further} results and new directions.
\newblock {\em Econometric reviews}, 26(1):53--90.

\bibitem[Green and Hastie, 2009]{green2009}
Green, P. and Hastie, D. (2009).
\newblock Reversible jump {{MCMC}}.
\newblock {\em Genetics}, 155.

\bibitem[Green, 1995]{green1995}
Green, P.~J. (1995).
\newblock Reversible {{Jump Markov Chain Monte Carlo Computation}} and
  {{Bayesian Model Determination}}.
\newblock {\em Biometrika}, 82(4):711--732.

\bibitem[Hannan, 1965]{hannan1965estimation}
Hannan, E.~J. (1965).
\newblock The estimation of relationships involving distributed lags.
\newblock {\em Econometrica: Journal of the Econometric Society}, pages
  206--224.

\bibitem[Holmes and Held, 2006]{holmes2006}
Holmes, C.~C. and Held, L. (2006).
\newblock Bayesian auxiliary variable models for binary and multinomial
  regression.
\newblock {\em Bayesian Analysis}, 1(1):145--168.

\bibitem[Ji et~al., 2012]{ji2012}
Ji, Y., Lin, N., and Zhang, B. (2012).
\newblock Model selection in binary and tobit quantile regression using the
  {{Gibbs}} sampler.
\newblock {\em Computational Statistics \& Data Analysis}, 56(4):827--839.

\bibitem[Johndrow et~al., 2018]{johndrow2018mcmc}
Johndrow, J.~E., Smith, A., Pillai, N., and Dunson, D.~B. (2018).
\newblock {MCMC} for imbalanced categorical data.
\newblock {\em Journal of the American Statistical Association}.

\bibitem[Kordas, 2006]{kordas2006}
Kordas, G. (2006).
\newblock Smoothed binary regression quantiles.
\newblock {\em Journal of Applied Econometrics}, 21(3):387--407.

\bibitem[Kozumi and Kobayashi, 2011]{kozumi2011gibbs}
Kozumi, H. and Kobayashi, G. (2011).
\newblock Gibbs sampling methods for {{Bayesian}} quantile regression.
\newblock {\em Journal of Statistical Computation and Simulation},
  81(11):1565--1578.

\bibitem[Liu and Wu, 1999]{liu1999parameter}
Liu, J.~S. and Wu, Y.~N. (1999).
\newblock Parameter expansion for data augmentation.
\newblock {\em Journal of the American Statistical Association},
  94(448):1264--1274.

\bibitem[L{\"u}tkepohl, 1981]{lutkepohl1981model}
L{\"u}tkepohl, H. (1981).
\newblock A model for non-negative and non-positive distributed lag functions.
\newblock {\em Journal of Econometrics}, 16(2):211--219.

\bibitem[Maheshwari and Rahman, 2023]{maheshwari2023bqror}
Maheshwari, P. and Rahman, M.~A. (2023).
\newblock bqror: an {R} package for {Bayesian} quantile regression in ordinal
  models.
\newblock {\em The R Journal}, 15:39--55.

\bibitem[Meier et~al., 2008]{meier2008group}
Meier, L., Van De~Geer, S., and B{\"u}hlmann, P. (2008).
\newblock The group lasso for logistic regression.
\newblock {\em Journal of the Royal Statistical Society: Series B (Statistical
  Methodology)}, 70(1):53--71.

\bibitem[Mitchell and Beauchamp, 1988]{mitchell1988bayesian}
Mitchell, T.~J. and Beauchamp, J.~J. (1988).
\newblock Bayesian variable selection in linear regression.
\newblock {\em Journal of the american statistical association},
  83(404):1023--1032.

\bibitem[Mogliani and Simoni, 2020]{mogliani2020}
Mogliani, M. and Simoni, A. (2020).
\newblock Bayesian {{MIDAS Penalized Regressions}}: {{Estimation}},
  {{Selection}}, and {{Prediction}}.
\newblock {\em arXiv:1903.08025 [econ]}.

\bibitem[Piironen and Vehtari, 2017]{piironen2017sparsity}
Piironen, J. and Vehtari, A. (2017).
\newblock {Sparsity information and regularization in the horseshoe and other
  shrinkage priors}.
\newblock {\em Electronic Journal of Statistics}, 11(2):5018 -- 5051.

\bibitem[Pillow and Scott, 2012]{pillow2012fully}
Pillow, J. and Scott, J. (2012).
\newblock Fully {Bayesian} inference for neural models with negative-binomial
  spiking.
\newblock {\em Advances in neural information processing systems}, 25.

\bibitem[Polson et~al., 2013]{polson2013}
Polson, N.~G., Scott, J.~G., and Windle, J. (2013).
\newblock Bayesian {{Inference}} for {{Logistic Models Using
  P\'olya}}\textendash{{Gamma Latent Variables}}.
\newblock {\em Journal of the American Statistical Association},
  108(504):1339--1349.

\bibitem[Rahman, 2016]{rahman2016bayesian}
Rahman, M.~A. (2016).
\newblock Bayesian quantile regression for ordinal models.
\newblock {\em Bayesian Analysis}, 11(1):1--24.

\bibitem[Rubin, 1986]{rubin1986statistical}
Rubin, D.~B. (1986).
\newblock Statistical matching using file concatenation with adjusted weights
  and multiple imputations.
\newblock {\em Journal of Business \& Economic Statistics}, 4(1):87--94.

\bibitem[Rushworth et~al., 2013]{rushworth2013distributed}
Rushworth, A.~M., Bowman, A.~W., Brewer, M.~J., and Langan, S.~J. (2013).
\newblock Distributed lag models for hydrological data.
\newblock {\em Biometrics}, 69(2):537--544.

\bibitem[Schwartz, 2000]{schwartz2000}
Schwartz, J. (2000).
\newblock The {{Distributed Lag}} between {{Air Pollution}} and {{Daily
  Deaths}}.
\newblock {\em Epidemiology}, 11(3):320--326.

\bibitem[Warren et~al., 2020]{warren2020spatially}
Warren, J.~L., Luben, T.~J., and Chang, H.~H. (2020).
\newblock A spatially varying distributed lag model with application to an air
  pollution and term low birth weight study.
\newblock {\em Journal of the Royal Statistical Society. Series C, Applied
  statistics}, 69(3):681.

\bibitem[Wilson et~al., 2017]{wilson2017bayesian}
Wilson, A., Chiu, Y.-H.~M., Hsu, H.-H.~L., Wright, R.~O., Wright, R.~J., and
  Coull, B.~A. (2017).
\newblock Bayesian distributed lag interaction models to identify perinatal
  windows of vulnerability in children’s health.
\newblock {\em Biostatistics}, 18(3):537--552.

\bibitem[Xu and Ghosh, 2015]{xu2015bayesian}
Xu, X. and Ghosh, M. (2015).
\newblock Bayesian variable selection and estimation for group lasso.
\newblock {\em Bayesian Analysis}, 10(4):909--936.

\bibitem[Yu and Zhang, 2005]{yu2005three}
Yu, K. and Zhang, J. (2005).
\newblock A {{Three-Parameter Asymmetric Laplace Distribution}} and {{Its
  Extension}}.
\newblock {\em Communications in Statistics - Theory and Methods},
  34(9-10):1867--1879.

\bibitem[Zanobetti et~al., 2002]{zanobetti2002temporal}
Zanobetti, A., Schwartz, J., Samoli, E., Gryparis, A., Touloumi, G., Atkinson,
  R., Le~Tertre, A., Bobros, J., Celko, M., Goren, A., et~al. (2002).
\newblock The temporal pattern of mortality responses to air pollution: a
  multicity assessment of mortality displacement.
\newblock {\em Epidemiology}, pages 87--93.

\bibitem[Zens et~al., 2023]{zens2023ultimate}
Zens, G., Fr{\"u}hwirth-Schnatter, S., and Wagner, H. (2023).
\newblock Ultimate {P{\'o}lya} {Gamma} {Samplers}--{Efficient} {MCMC} for
  possibly imbalanced binary and categorical data.
\newblock {\em arXiv preprint arXiv:2011.06898}.

\bibitem[Zhou et~al., 2012]{zhou2012lognormal}
Zhou, M., Li, L., Dunson, D., and Carin, L. (2012).
\newblock Lognormal and gamma mixed negative binomial regression.
\newblock In {\em Proceedings of the... International Conference on Machine
  Learning. International Conference on Machine Learning}, volume 2012, page
  1343. NIH Public Access.

\end{thebibliography}
	
\end{document}


\begin{frontmatter}
\title{Bayesian Generalized Distributed Lag Regression with Variable Selection: Full Simulation Results Supplemental}
\runtitle{Full Simulation Results Supplemental}

\begin{aug}
\author[A]{\fnms{Daniel}~\snm{Dempsey}\ead[label=e1]{first@somewhere.com}},
\and
\author[B]{\fnms{Jason}~\snm{Wyse}\ead[label=e3]{third@somewhere.com}}
%
\end{aug}
%
%

\end{frontmatter}

\subsection*{Introduction}
In Section 3 of the manuscript that this file supplements, we performed simulation studies for both negative binomial DLMs and quantile binary DLMs. For each model, we created four sets of `large' datasets ($N = 5114$) and four sets of `small' datasets ($N = 250$). Further details around parameter specification and how the lag-response was simulated is given in Section 3.

In the main manuscript, to save space, we only reported results for the first large and small simulated datasets for both regression models. The purpose of this supplemental file is to provide \emph{all} the results for transparency. The visualisations given here follow the same format as the ones given in the main manuscript - that is, they report the agreement between the fitted model parameters and the true simulation values. The first set of graphs are results for negative binomial DLMs, followed by binary quantile DLMs, and the captions describe the parameter being visualised.

\newpage 
\subsection*{Negative Binomial Simulation Results}

\begin{myfig}
	\centering
	\includegraphics[scale=0.4]{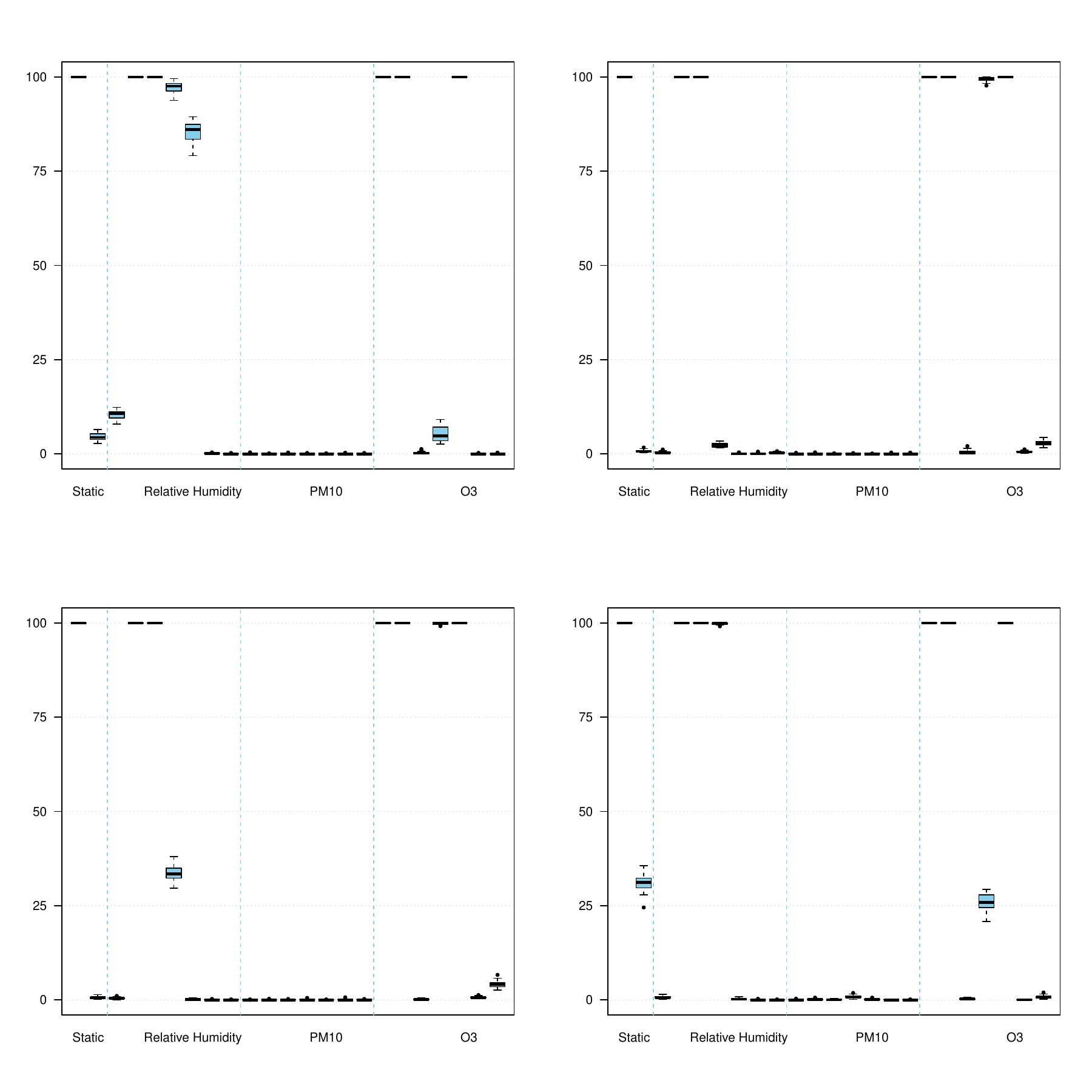}
	\captionof{figure}{Covariate inclusion parameters for large simulations.}
\end{myfig}

\newpage

\begin{myfig}
	\centering
	\includegraphics[scale=0.4]{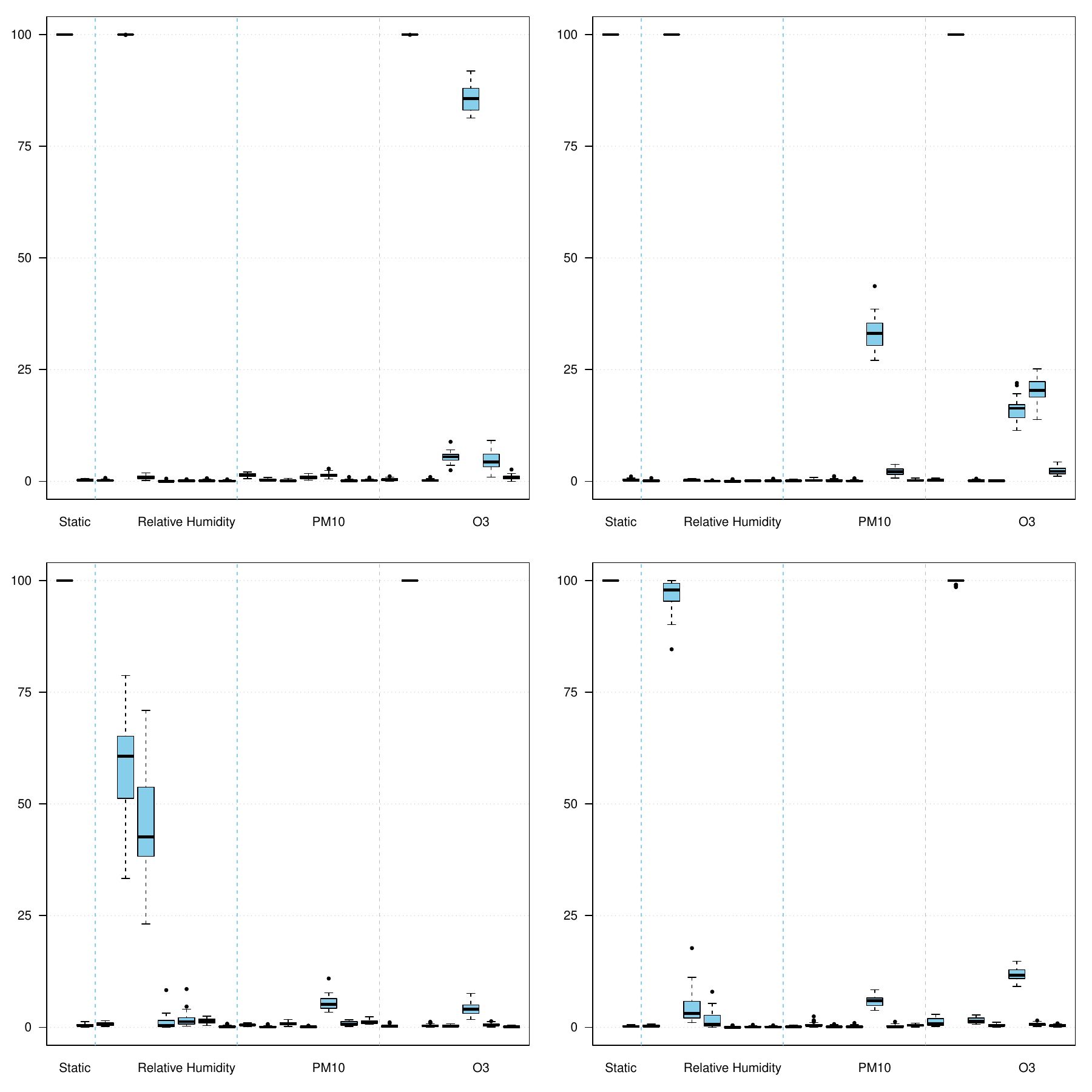}
	\captionof{figure}{Covariate inclusion parameters for small simulations.}
\end{myfig}

\newpage

\begin{myfig}
	\centering
	\includegraphics[scale=0.45]{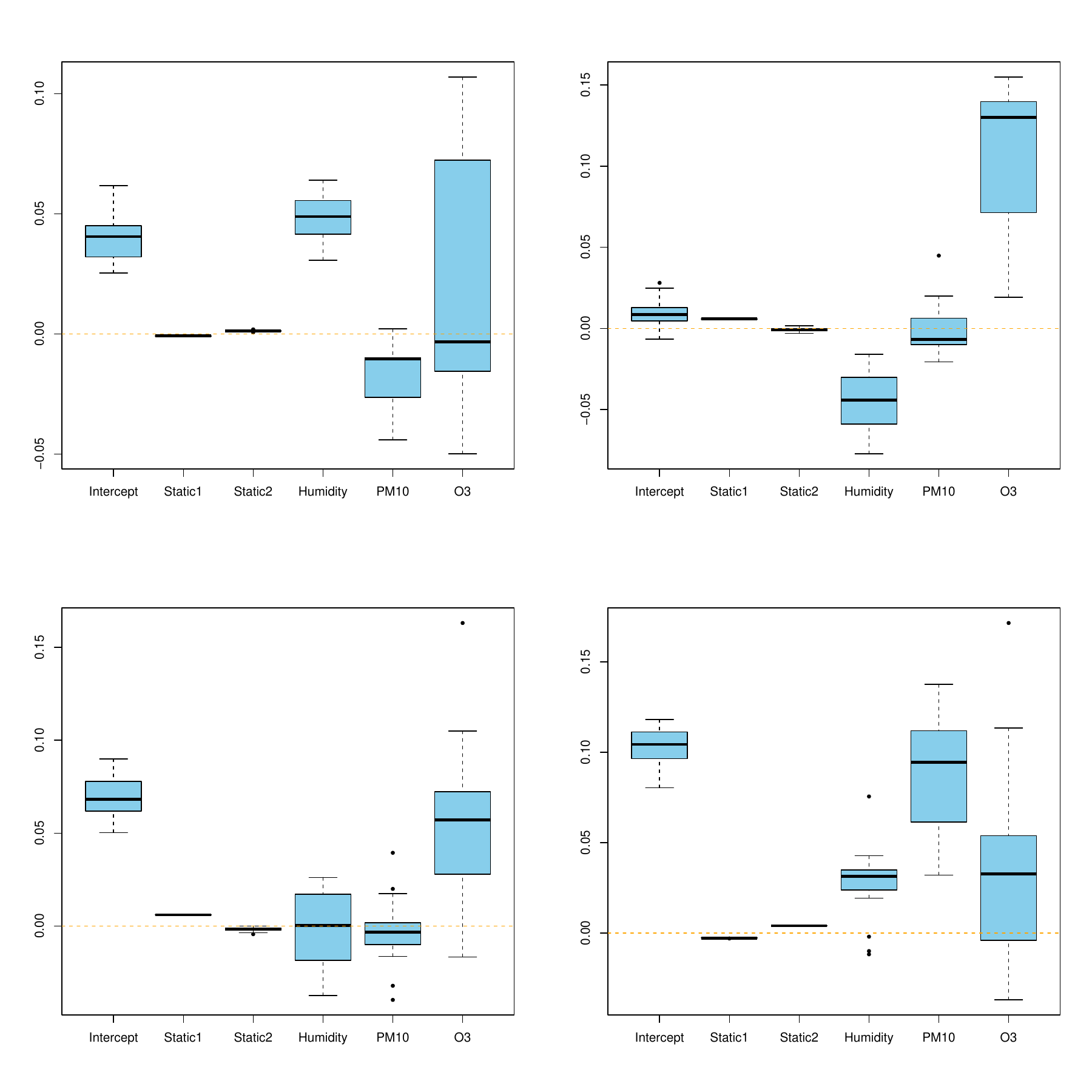}
	\captionof{figure}{Slope parameters for large simulations.}
\end{myfig}

\begin{myfig}
	\centering
	\includegraphics[scale=0.45]{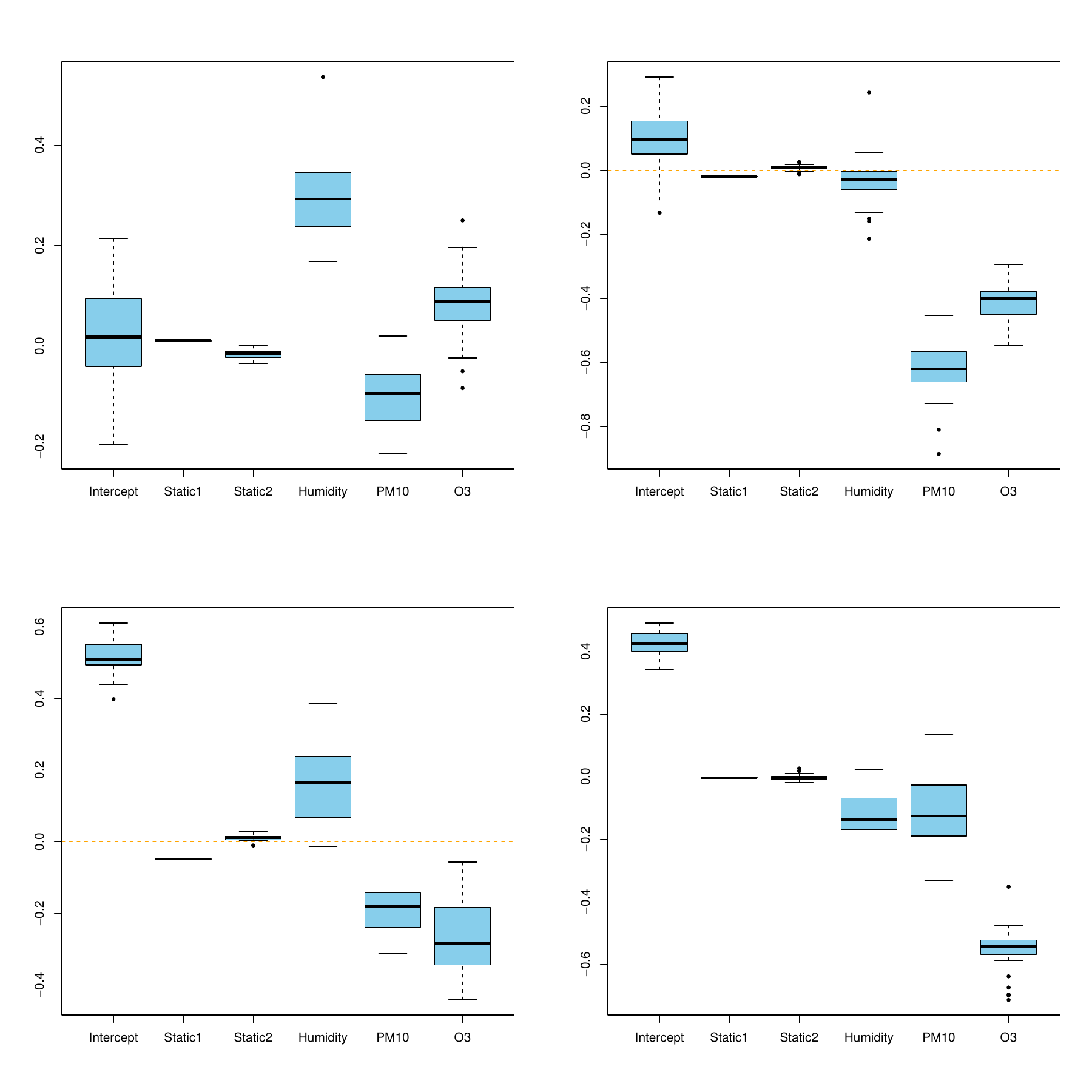}
	\captionof{figure}{Slope parameters for small simulations.}
\end{myfig}

\newpage
\setcounter{figure}{3}

\begin{figure}[hbt!]
	\centering
	\begin{subfigure}[b]{35em}
		\centering
		\includegraphics[scale=0.5]{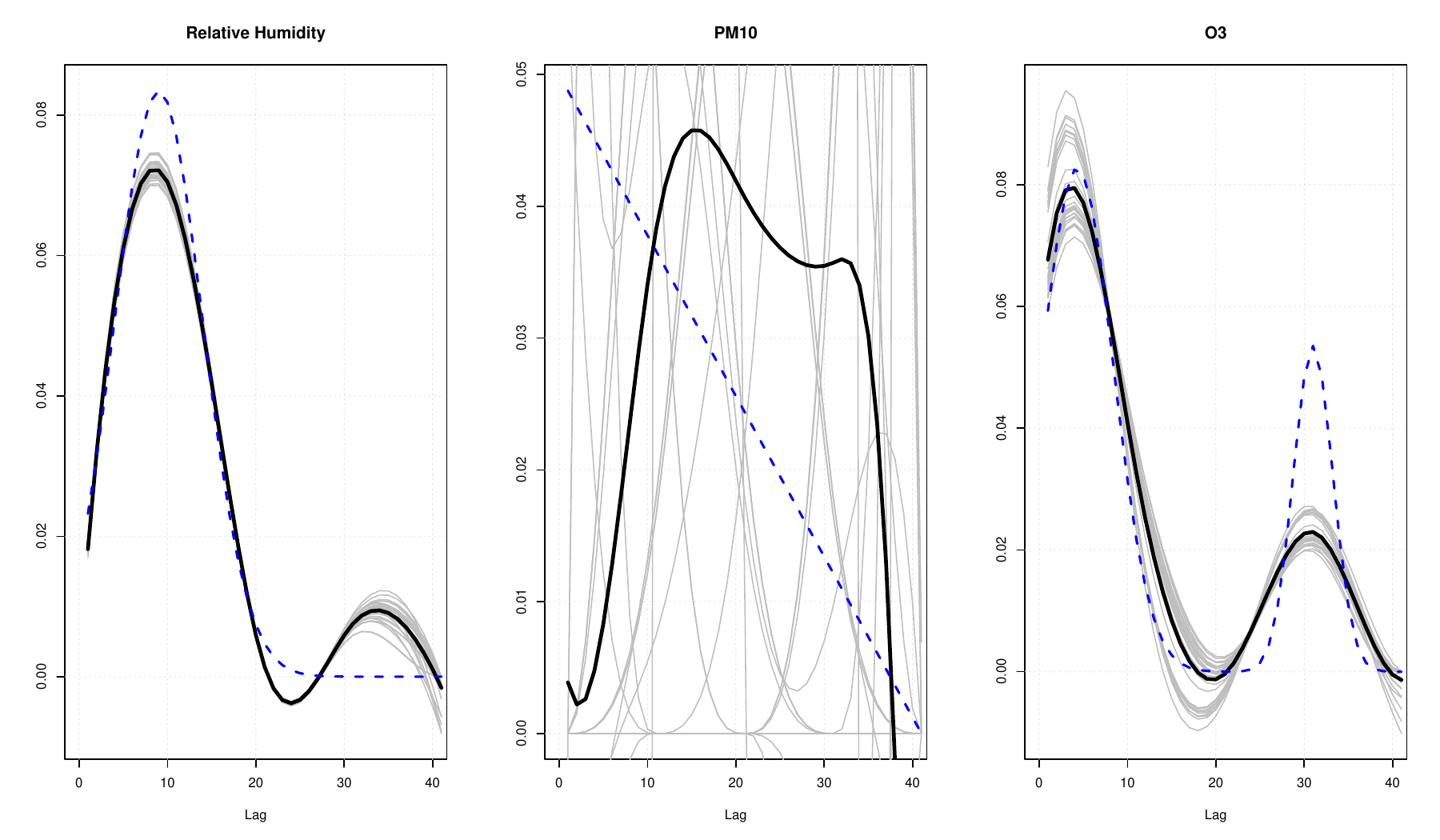}
	\end{subfigure}
	\hfill
	\begin{subfigure}[b]{37em}
		\centering
		\includegraphics[scale=0.5]{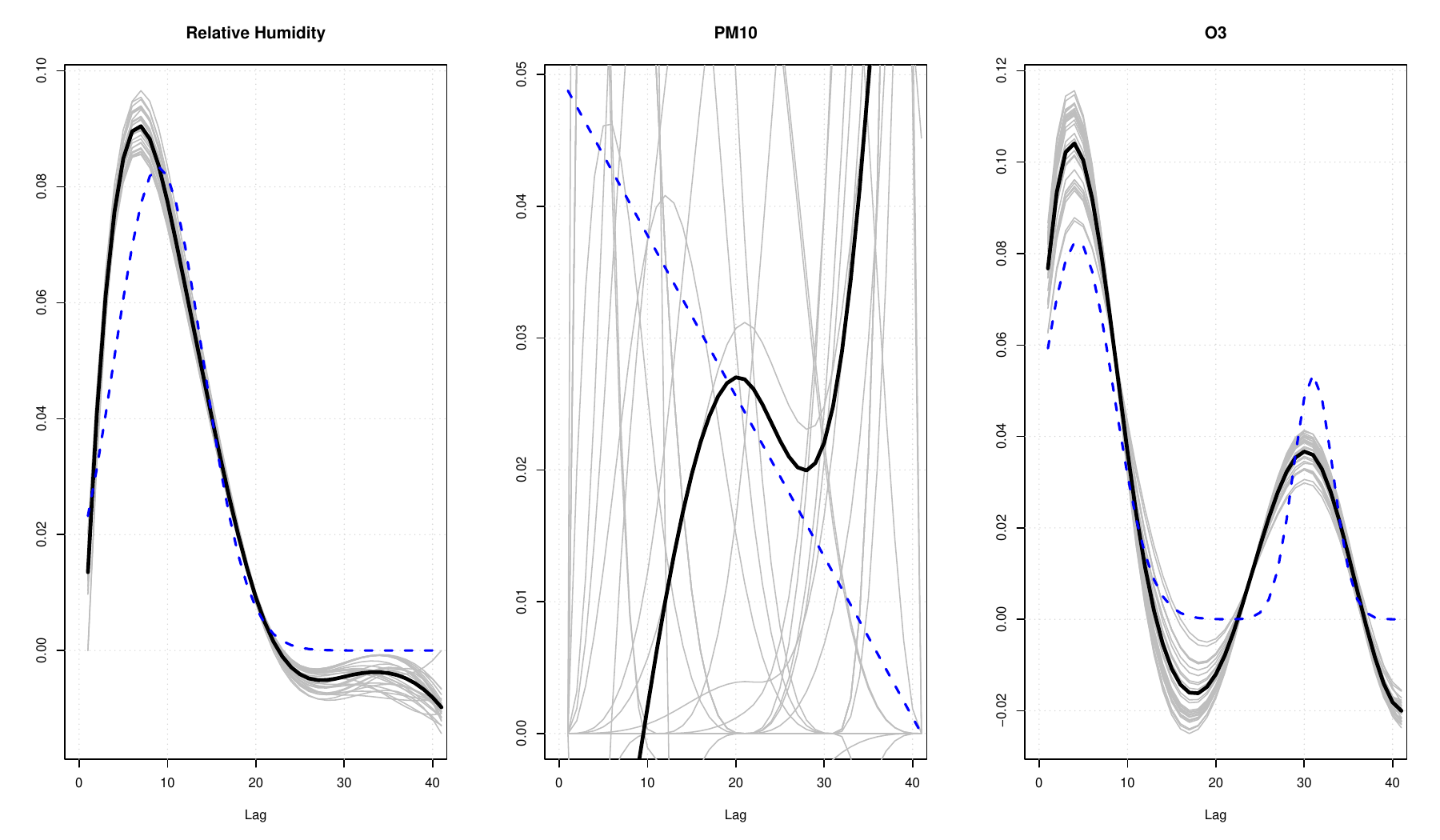}
	\end{subfigure}
	\captionof{figure}{Estimated lag-response for the first and second large simulations.}
\end{figure}

\newpage
\setcounter{figure}{4}

\begin{figure}[hbt!]
	\centering
	\begin{subfigure}[b]{37em}
		\centering
		\includegraphics[scale=0.5]{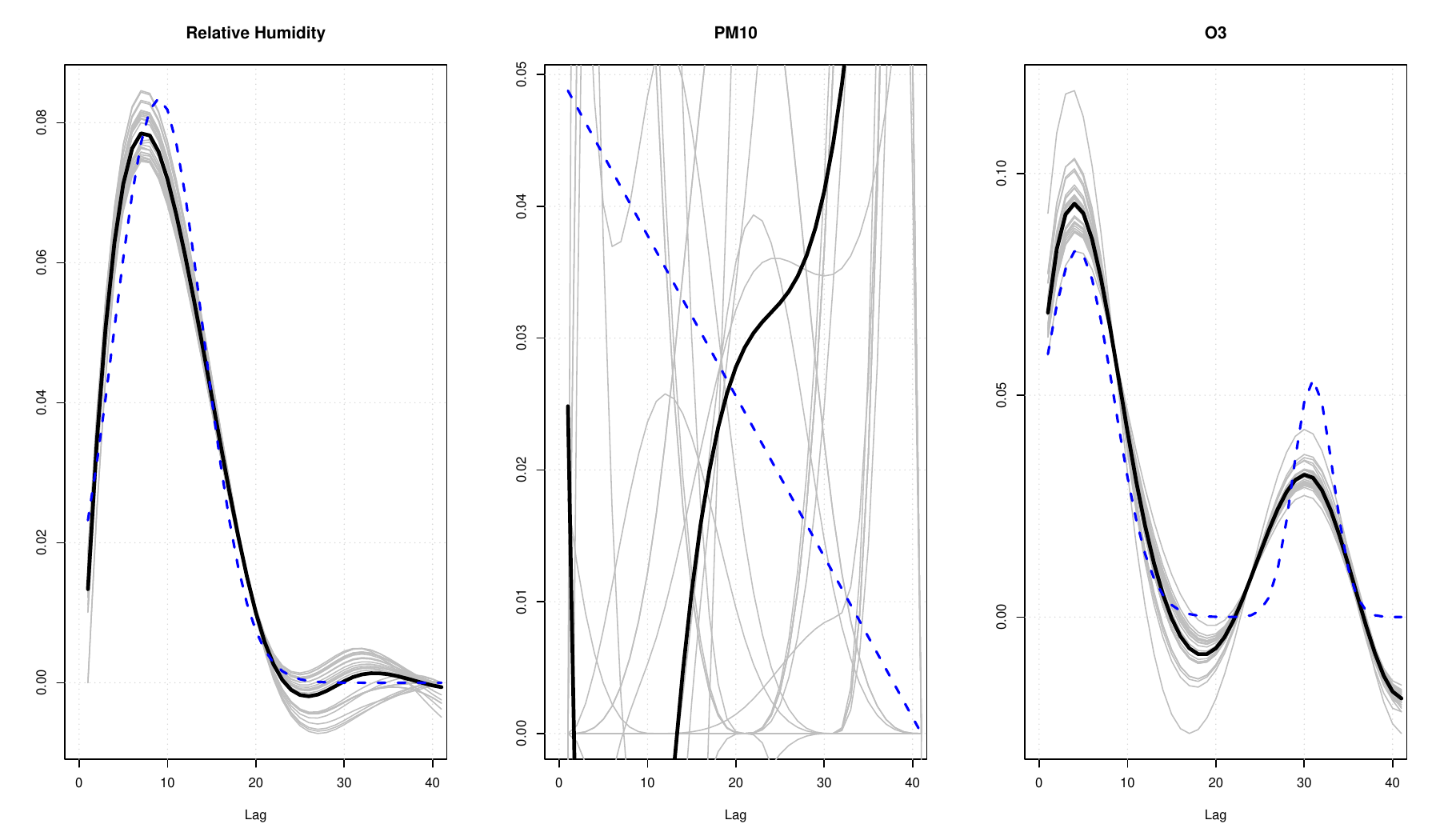}
	\end{subfigure}
	\hfill
	\begin{subfigure}[b]{37em}
		\centering
		\includegraphics[scale=0.5]{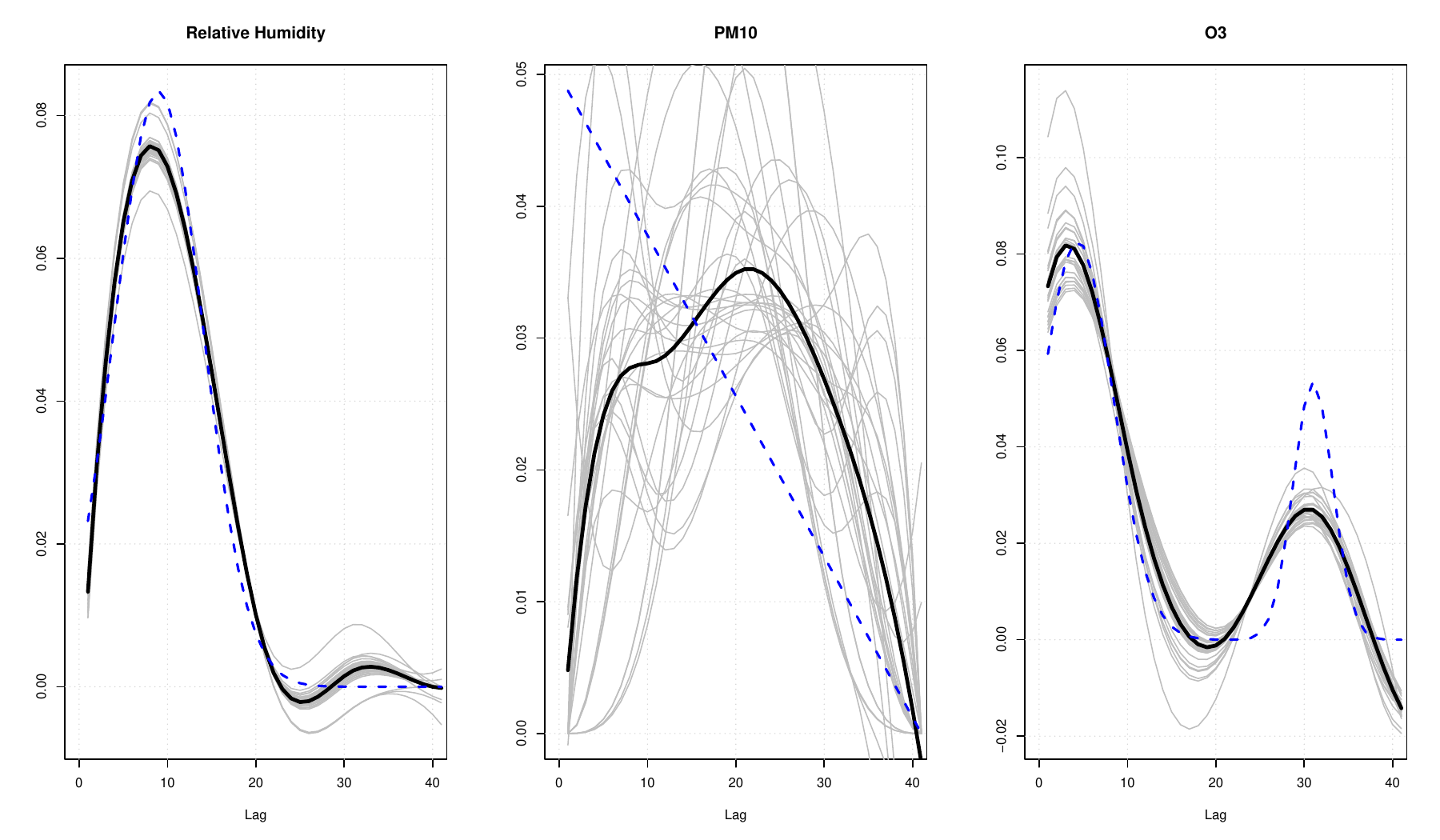}
	\end{subfigure}
	\captionof{figure}{Estimated lag-response for the third and fourth large simulations.}
\end{figure}

\newpage
\setcounter{figure}{5}

\begin{figure}[hbt!]
	\centering
	\begin{subfigure}[b]{37em}
		\centering
		\includegraphics[scale=0.5]{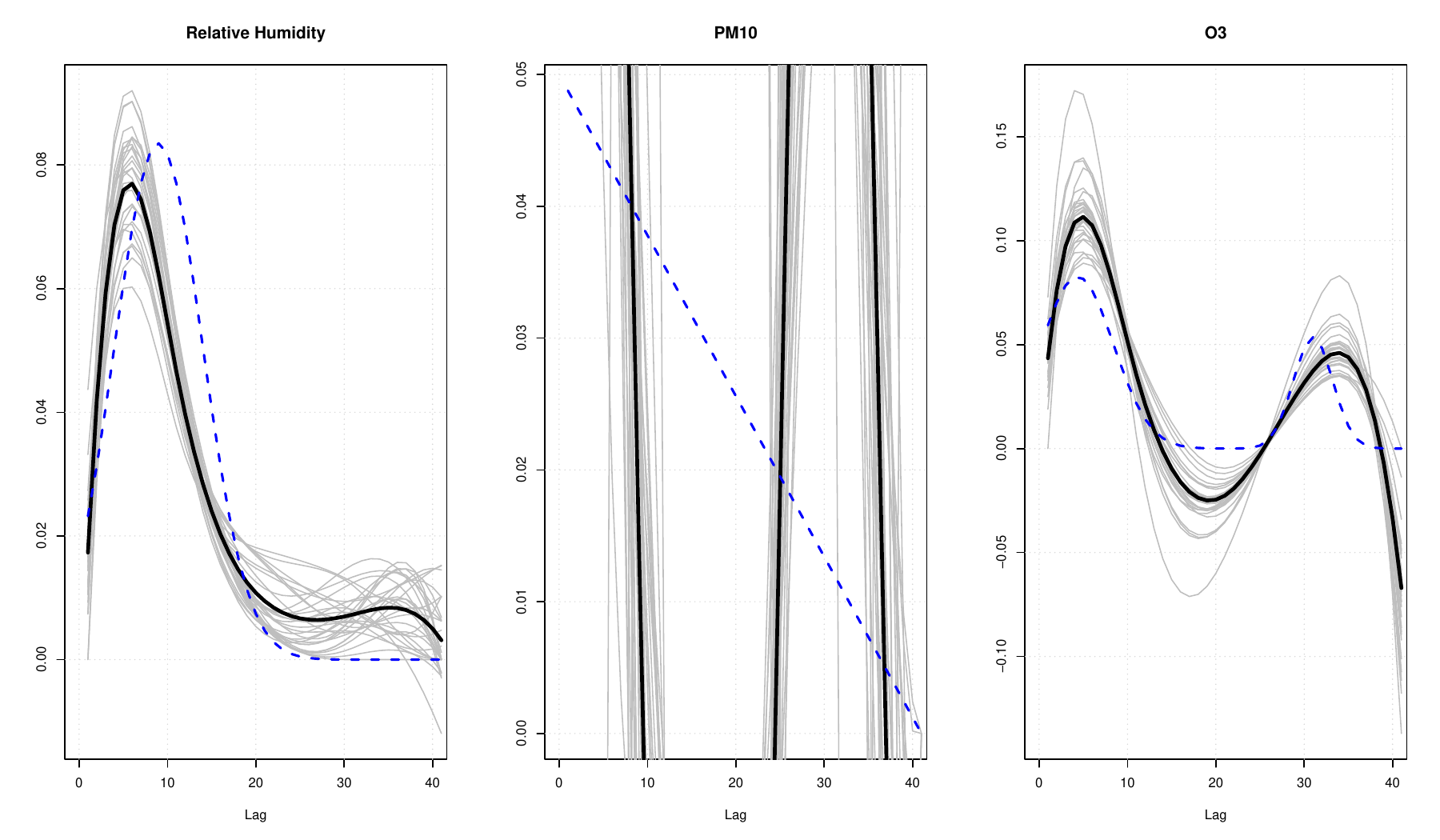}
	\end{subfigure}
	\hfill
	\begin{subfigure}[b]{37em}
		\centering
		\includegraphics[scale=0.5]{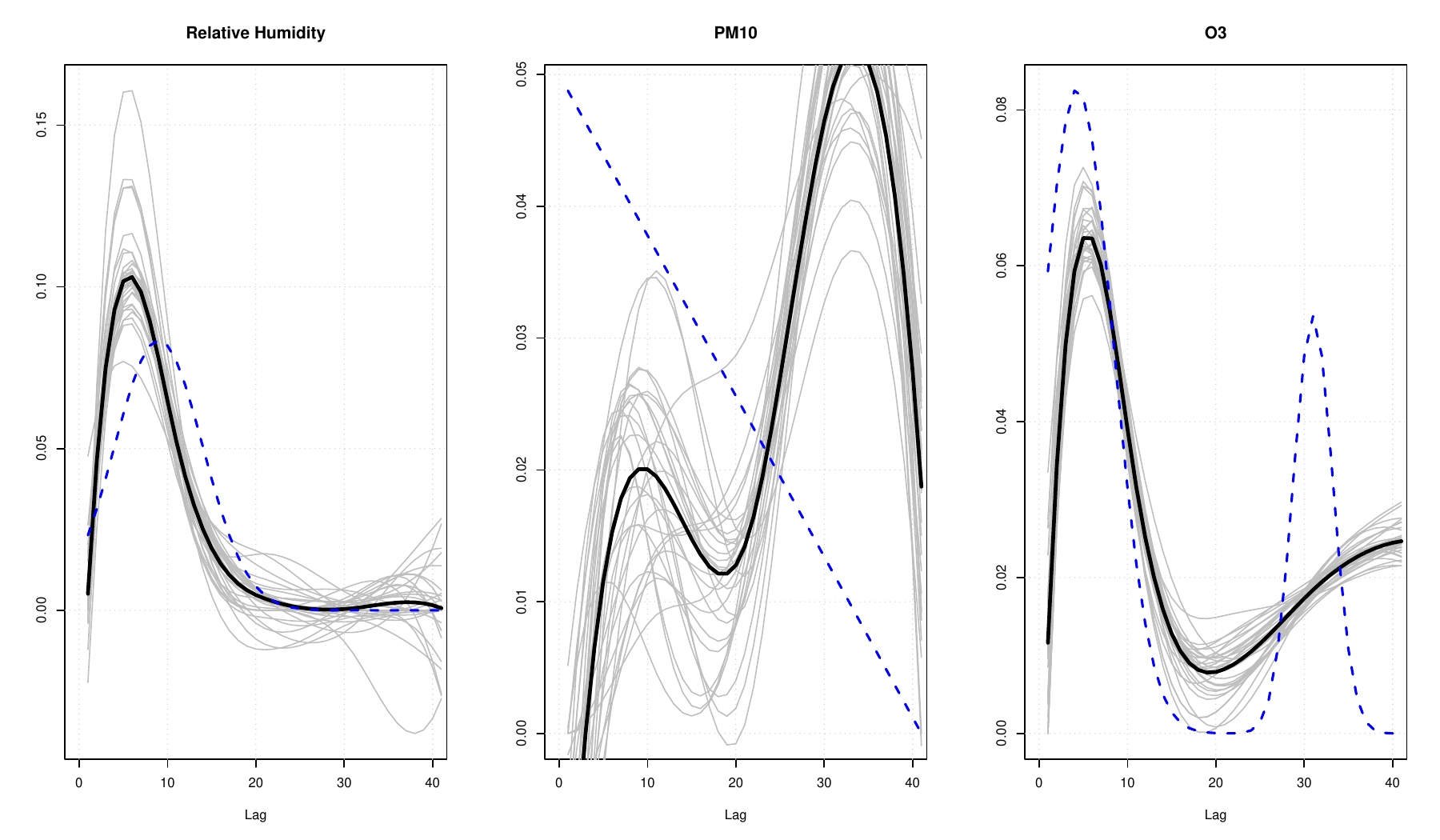}
	\end{subfigure}
	\captionof{figure}{Estimated lag-response for the first and second small simulations.}
\end{figure}

\newpage
\setcounter{figure}{6}

\begin{figure}[hbt!]
	\centering
	\begin{subfigure}[b]{37em}
		\centering
		\includegraphics[scale=0.5]{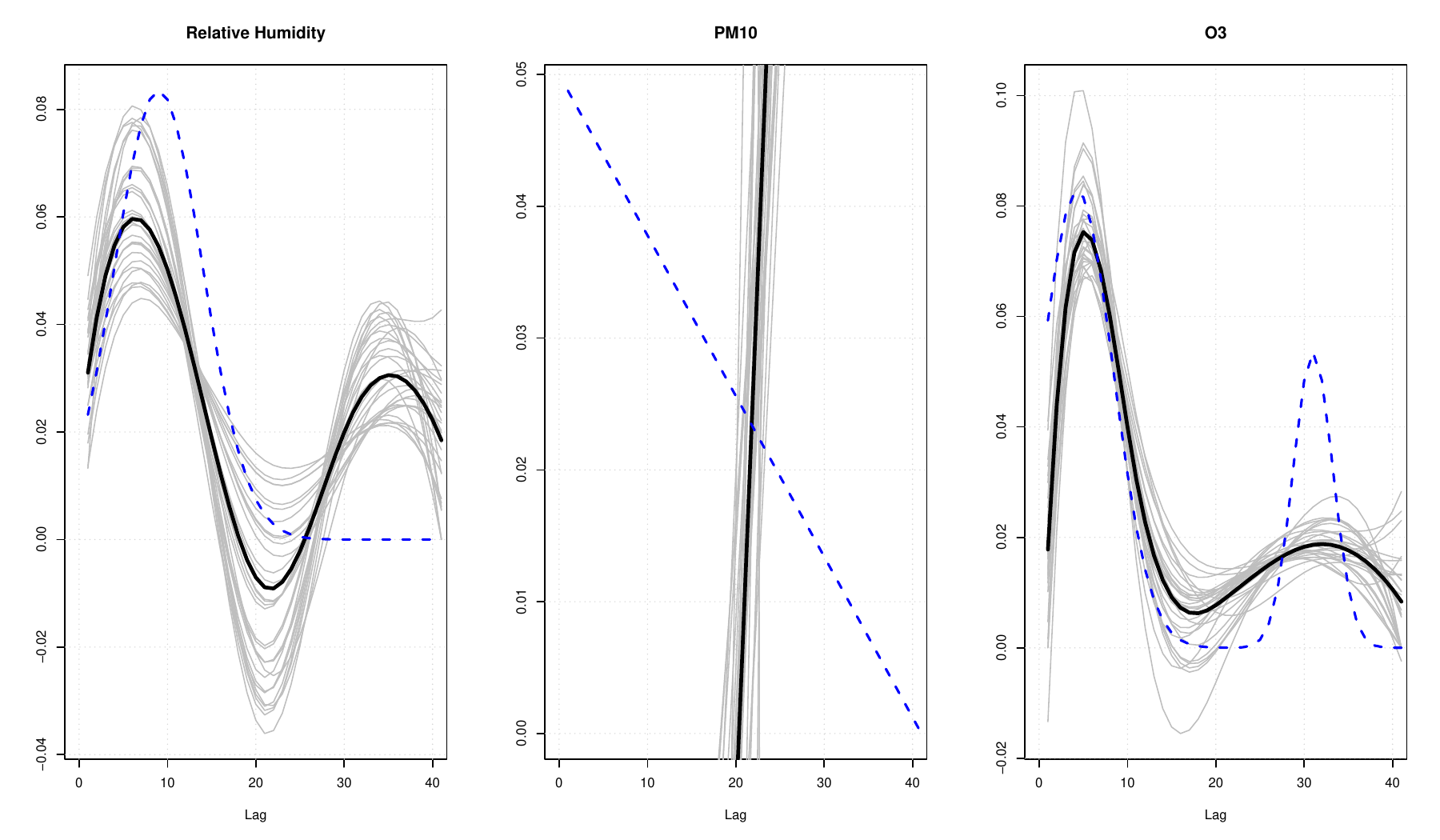}
	\end{subfigure}
	\hfill
	\begin{subfigure}[b]{37em}
		\centering
		\includegraphics[scale=0.5]{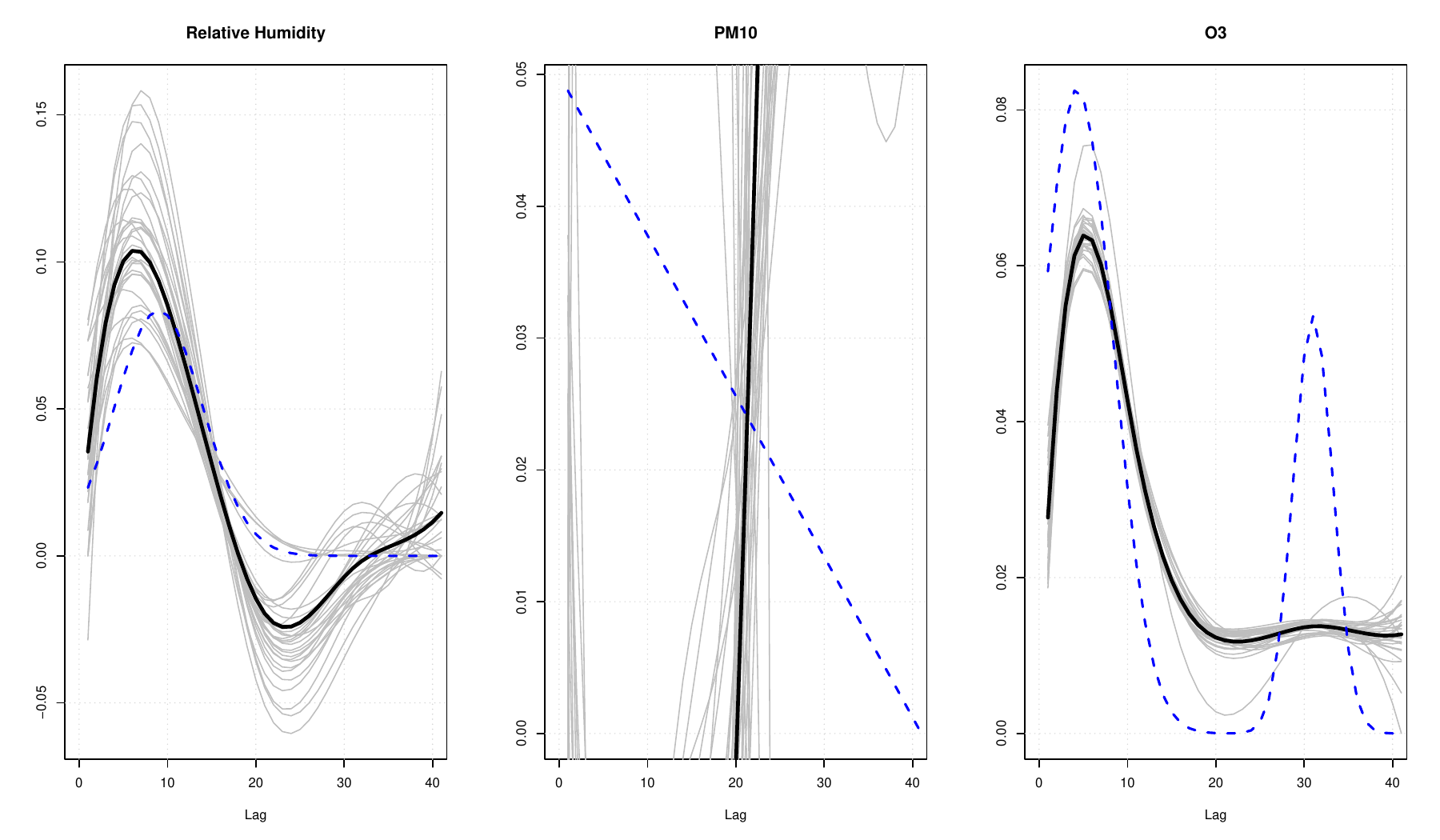}
	\end{subfigure}
	\captionof{figure}{Estimated lag-response for the third and fourth small simulations.}
\end{figure}

\newpage

\begin{myfig}
	\centering
	\includegraphics[scale=0.5]{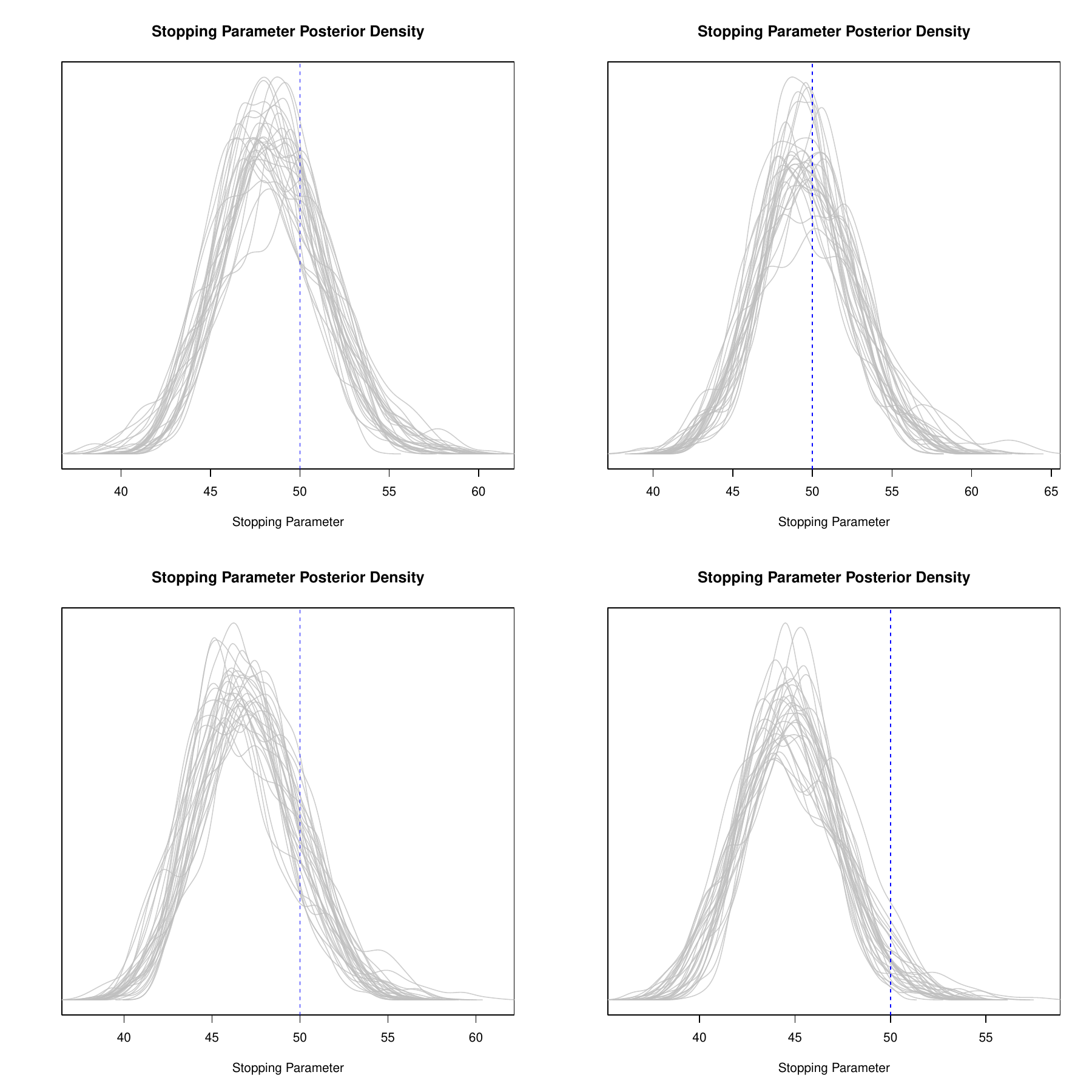}
	\captionof{figure}{Stopping parameter posteriors for large simulations.}
\end{myfig}

\begin{myfig}
	\centering
	\includegraphics[scale=0.5]{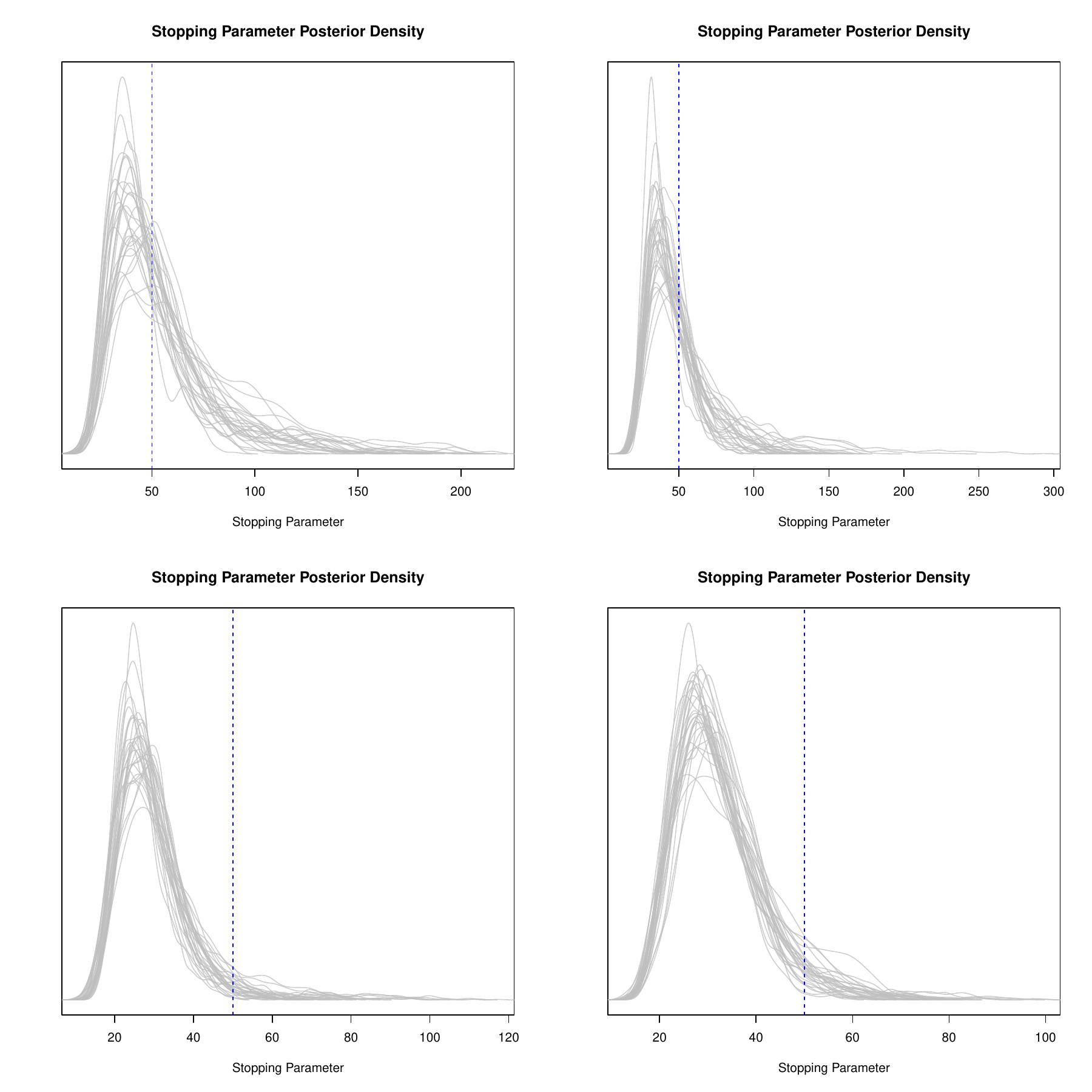}
	\captionof{figure}{Stopping parameter posteriors for small simulations.}
\end{myfig}

\newpage 

\subsection*{Quantile Binary Simulation Results}

\begin{myfig}
	\centering
	\includegraphics[scale=0.4]{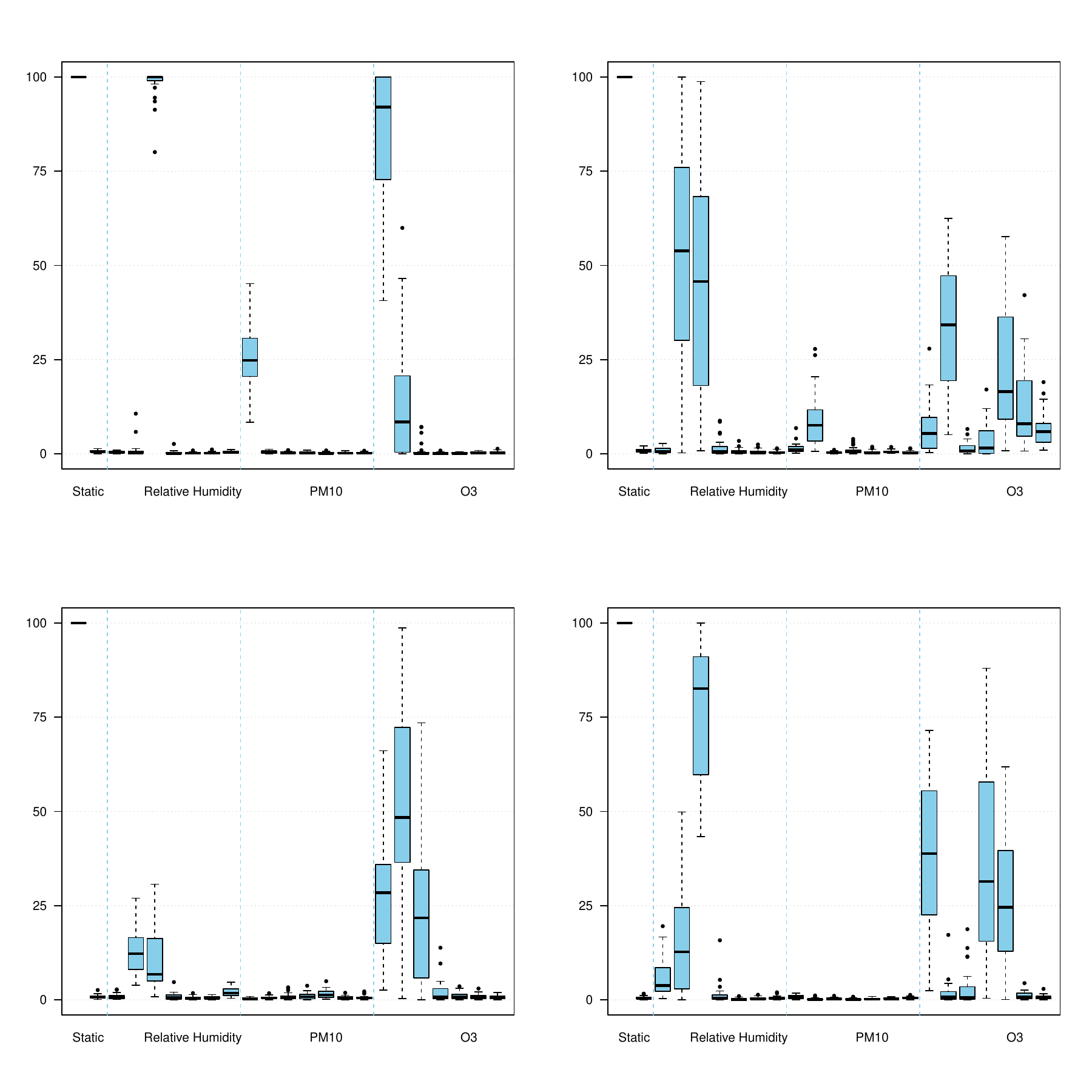}
	\captionof{figure}{Covariate inclusion parameters for large simulations.}
\end{myfig}

\begin{myfig}
	\centering
	\includegraphics[scale=0.4]{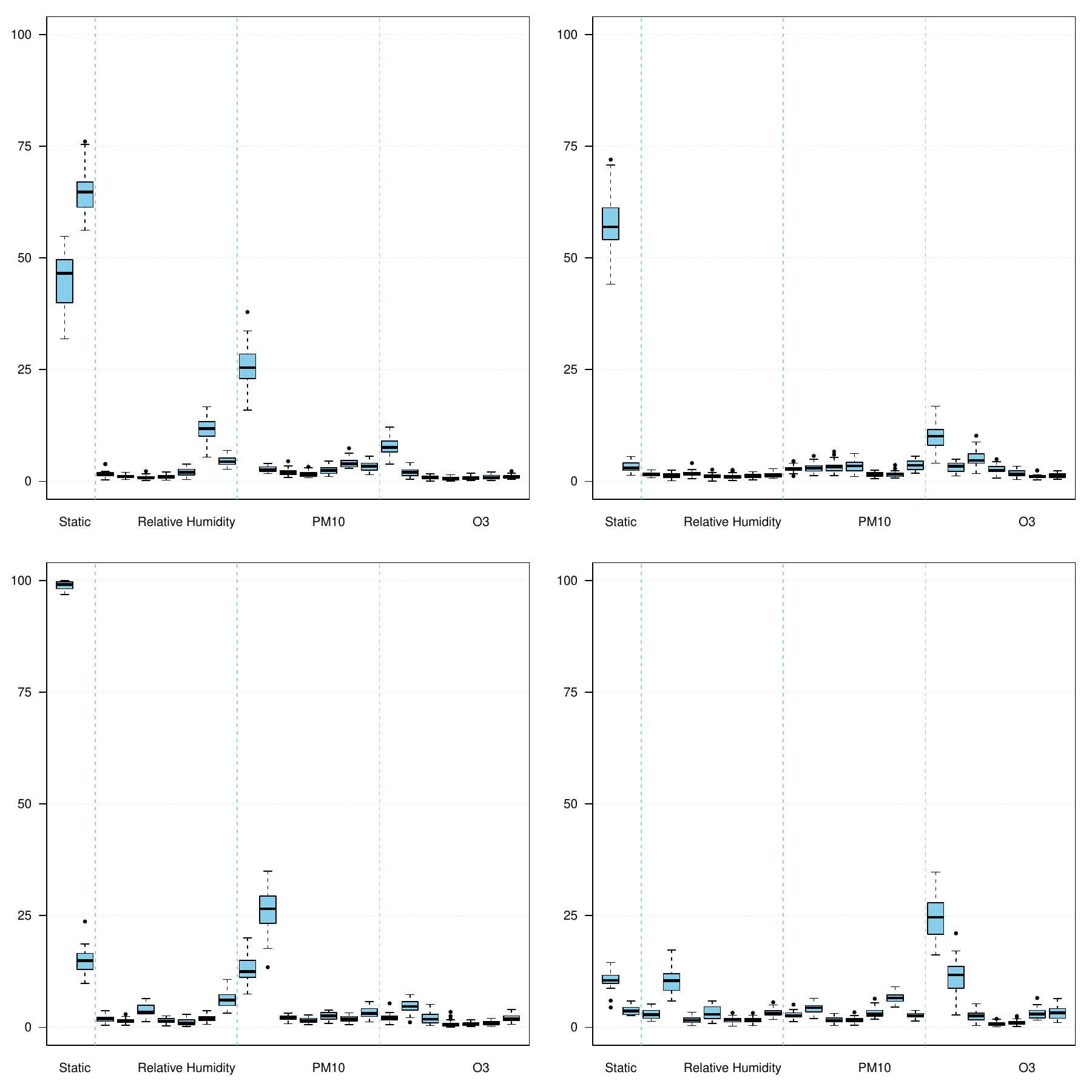}
	\captionof{figure}{Covariate inclusion parameters for small simulations.}
\end{myfig}

\begin{myfig}
	\centering
	\includegraphics[scale=0.45]{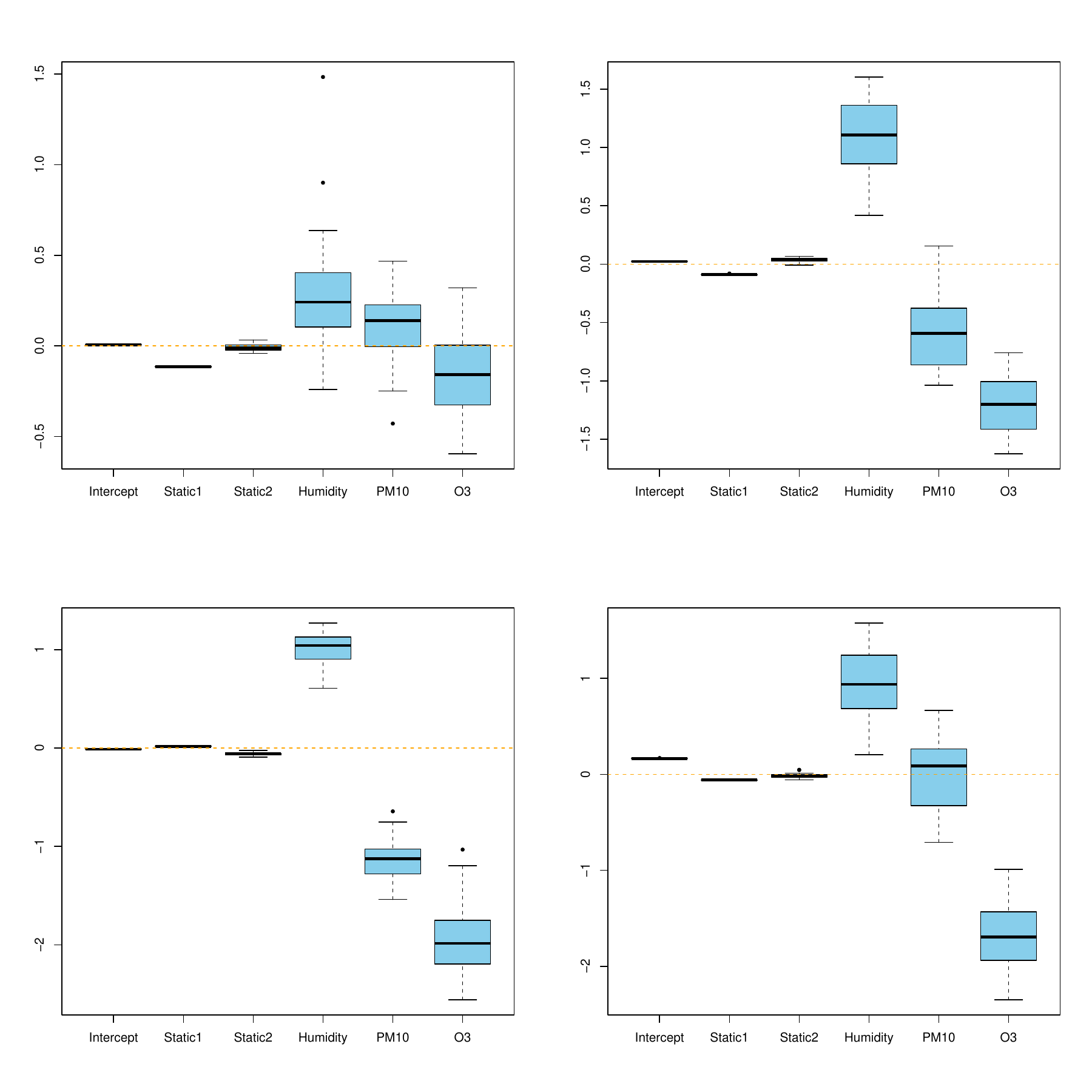}
	\captionof{figure}{Slope parameters for large simulations.}
\end{myfig}

\begin{myfig}
	\centering
	\includegraphics[scale=0.45]{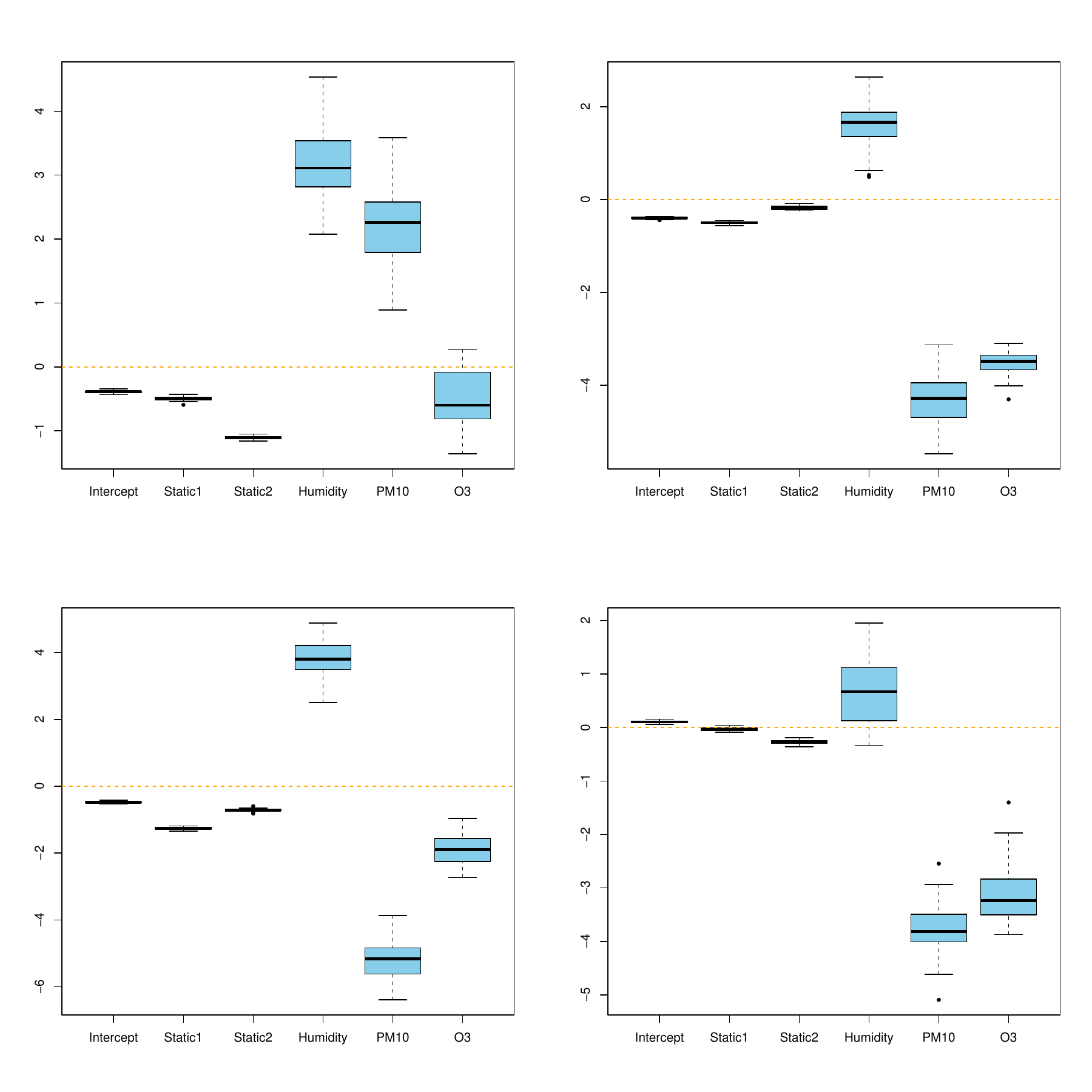}
	\captionof{figure}{Slope parameters for small simulations.}
\end{myfig}

\setcounter{figure}{13}
\begin{figure}[hbt!]
	\centering
	\begin{subfigure}[b]{37em}
		\centering
		\includegraphics[scale=0.5]{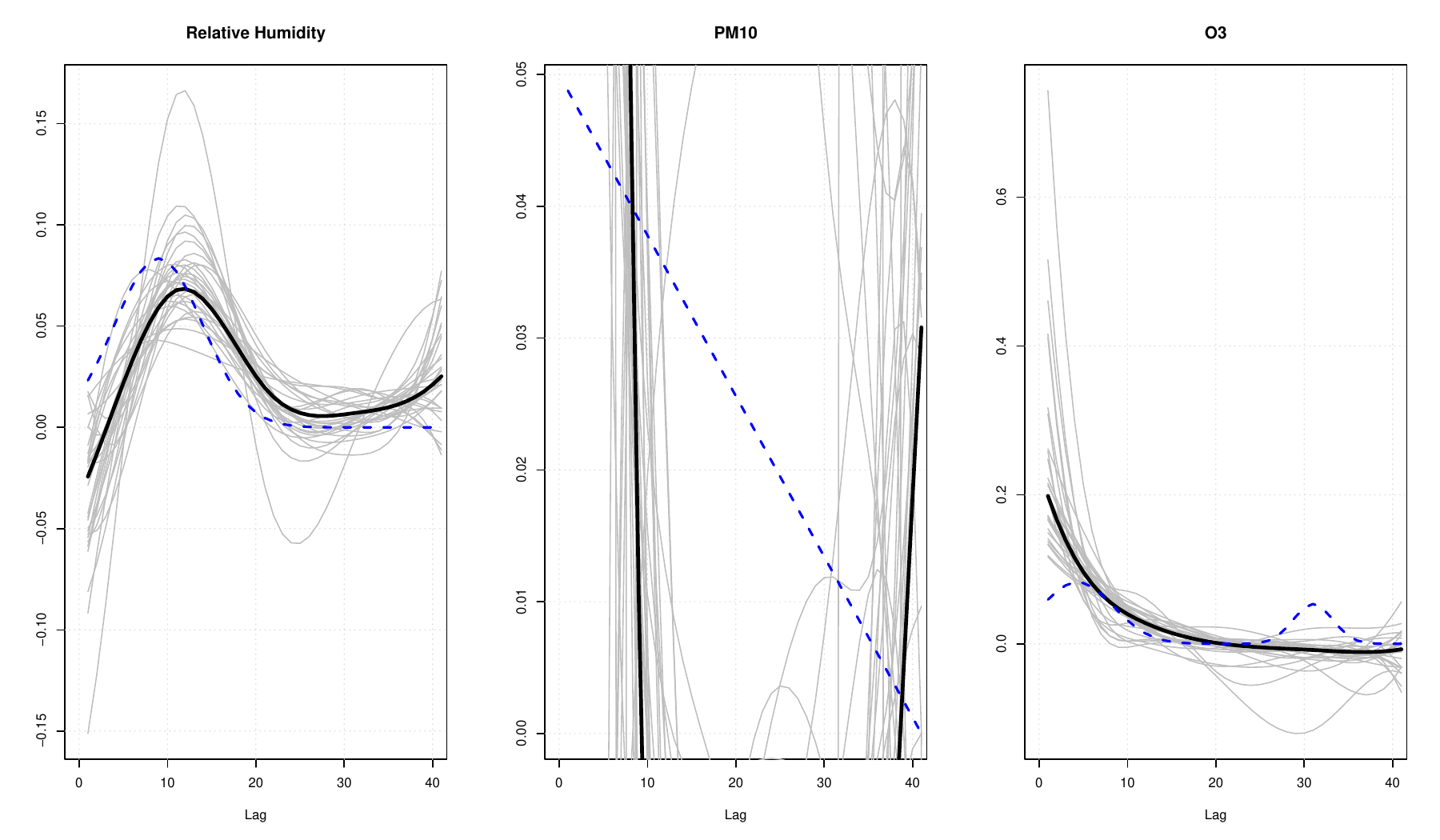}
	\end{subfigure}
	\hfill
	\begin{subfigure}[b]{37em}
		\centering
		\includegraphics[scale=0.5]{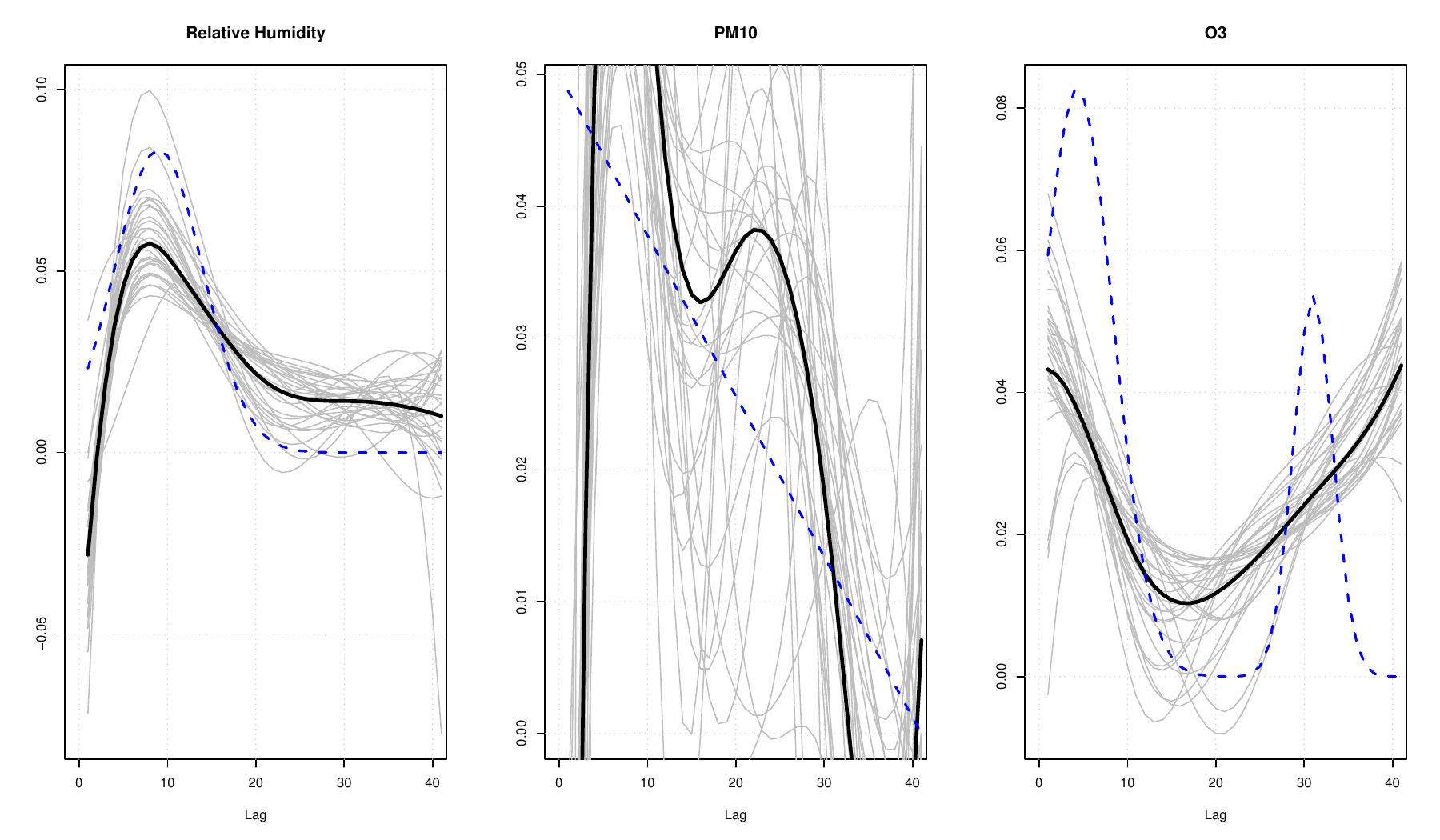}
	\end{subfigure}
	\captionof{figure}{Estimated lag-response for the first and second large simulations.}
\end{figure}

\setcounter{figure}{14}
\begin{figure}[hbt!]
	\centering
	\begin{subfigure}[b]{37em}
		\centering
		\includegraphics[scale=0.5]{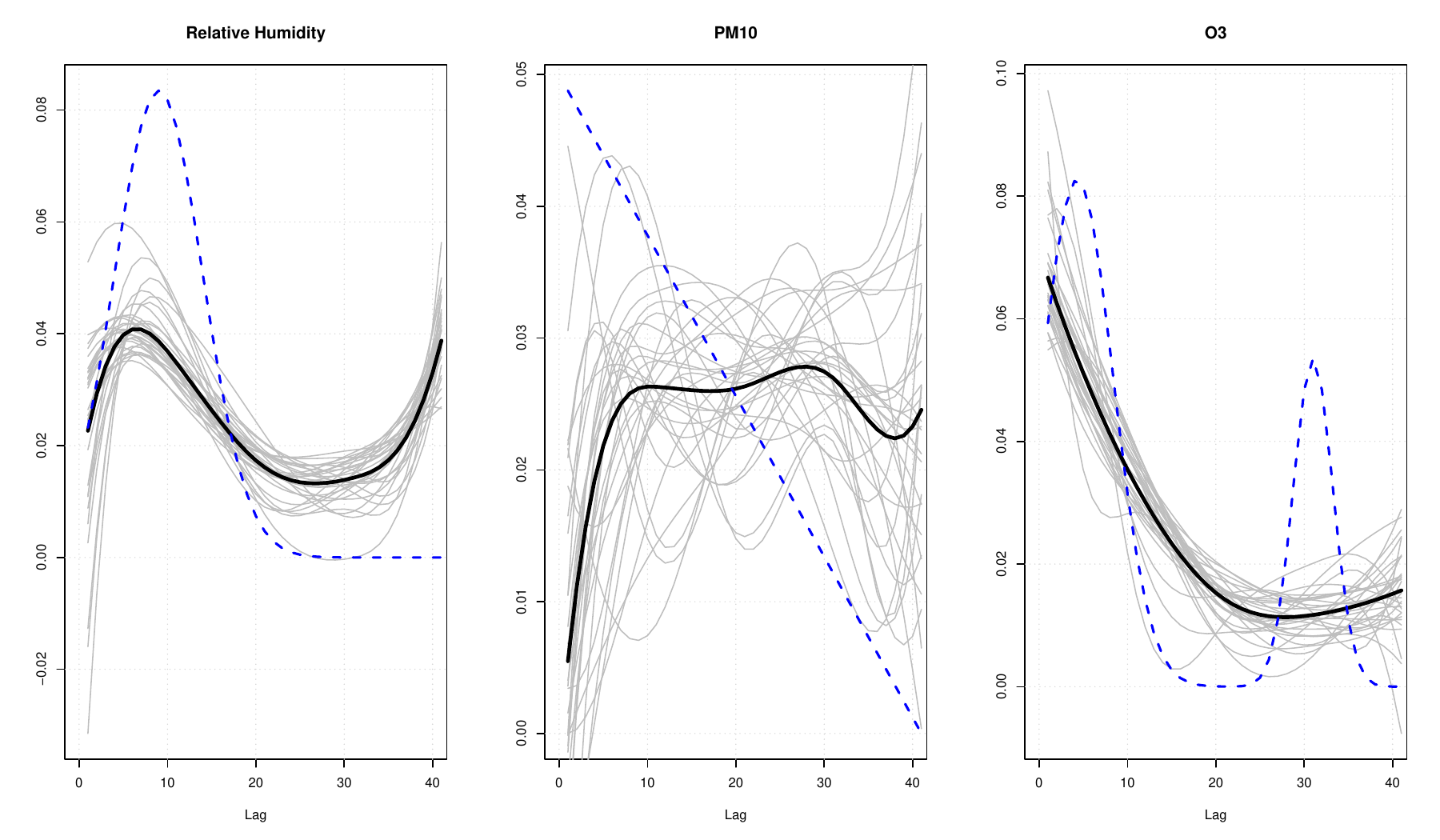}
	\end{subfigure}
	\hfill
	\begin{subfigure}[b]{37em}
		\centering
		\includegraphics[scale=0.5]{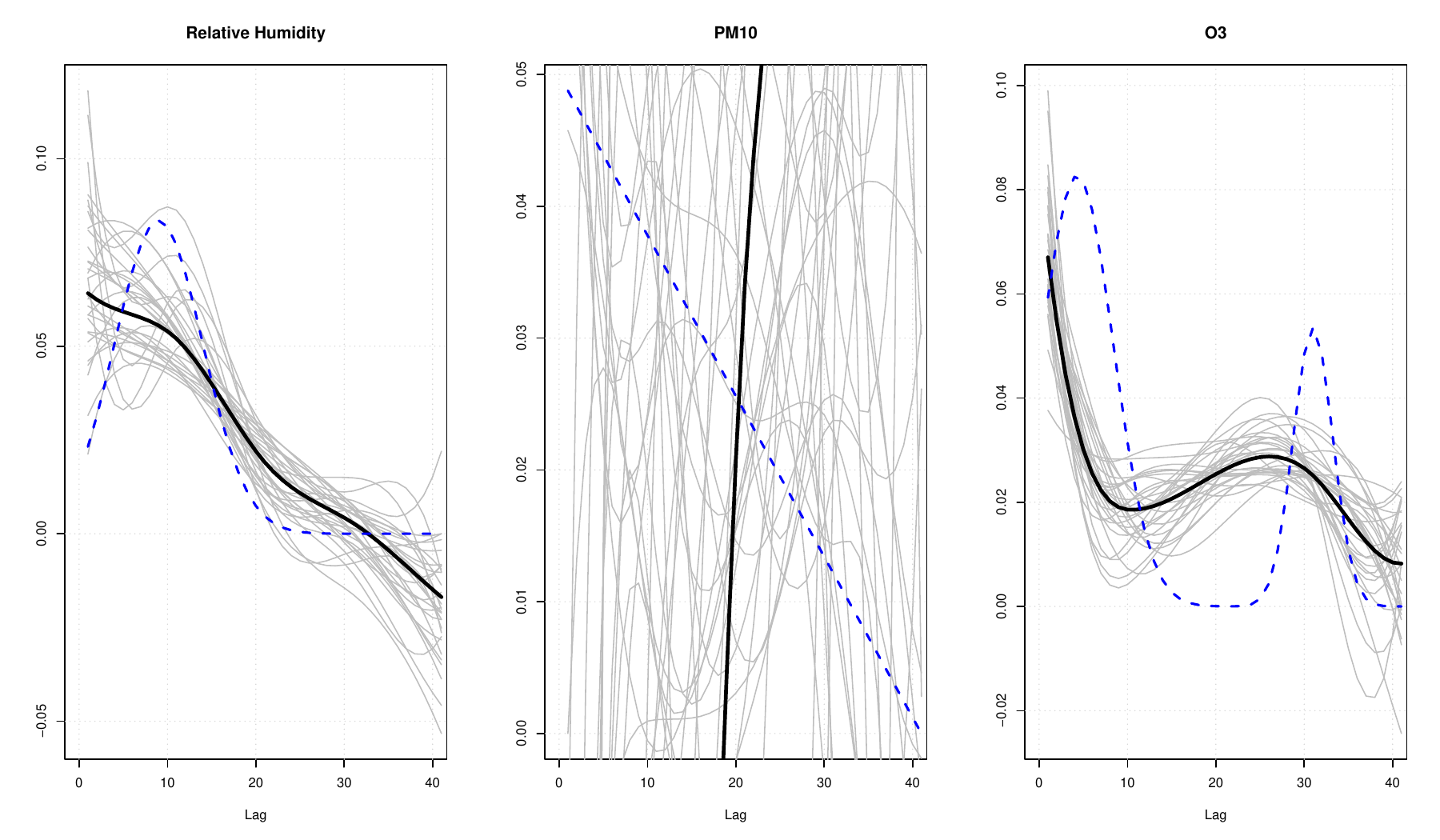}
	\end{subfigure}
	\captionof{figure}{Estimated lag-response for the third and fourth large simulations.}
\end{figure}

\setcounter{figure}{15}
\begin{figure}[hbt!]
	\centering
	\begin{subfigure}[b]{37em}
		\centering
		\includegraphics[scale=0.5]{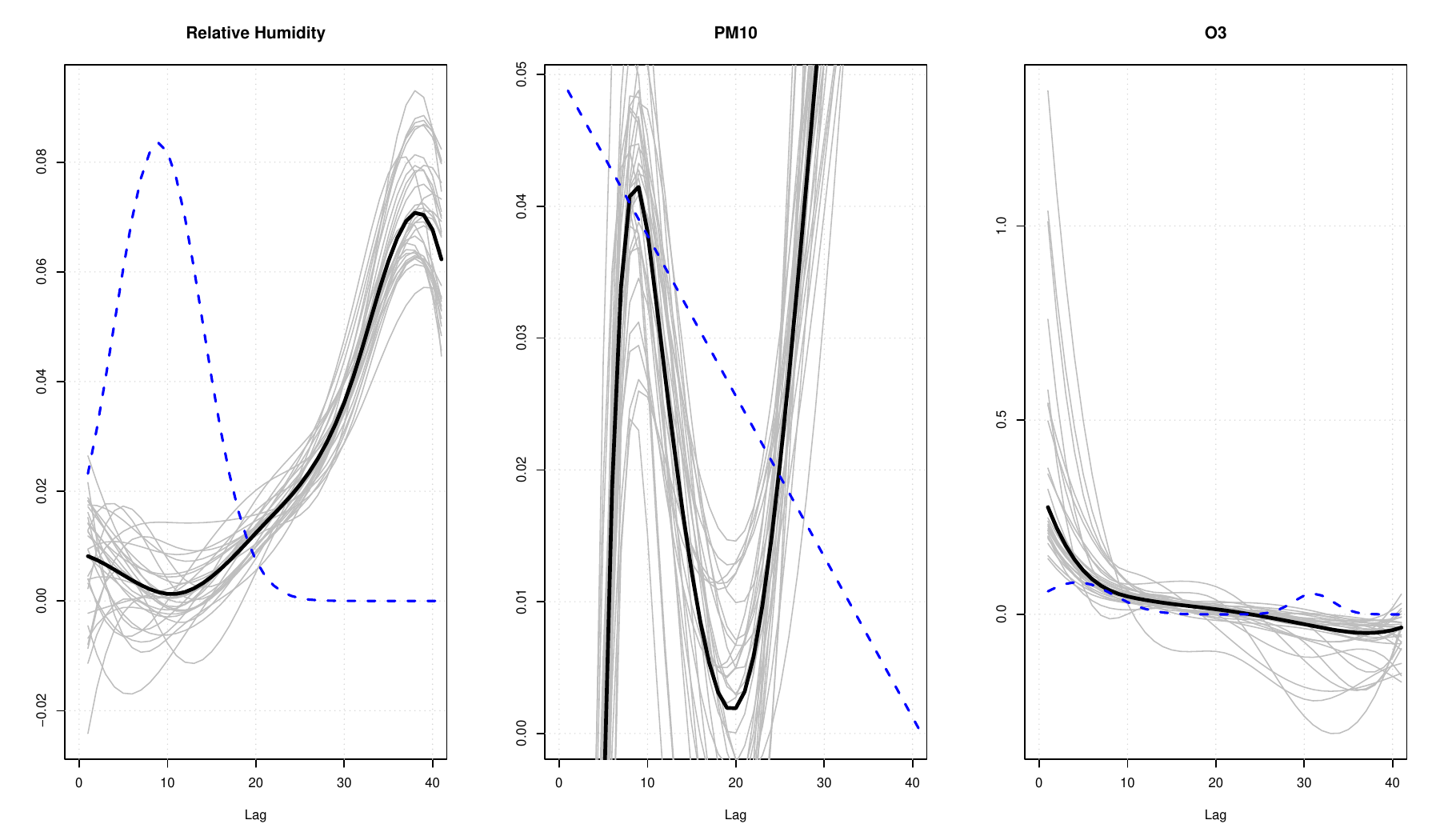}
	\end{subfigure}
	\hfill
	\begin{subfigure}[b]{37em}
		\centering
		\includegraphics[scale=0.5]{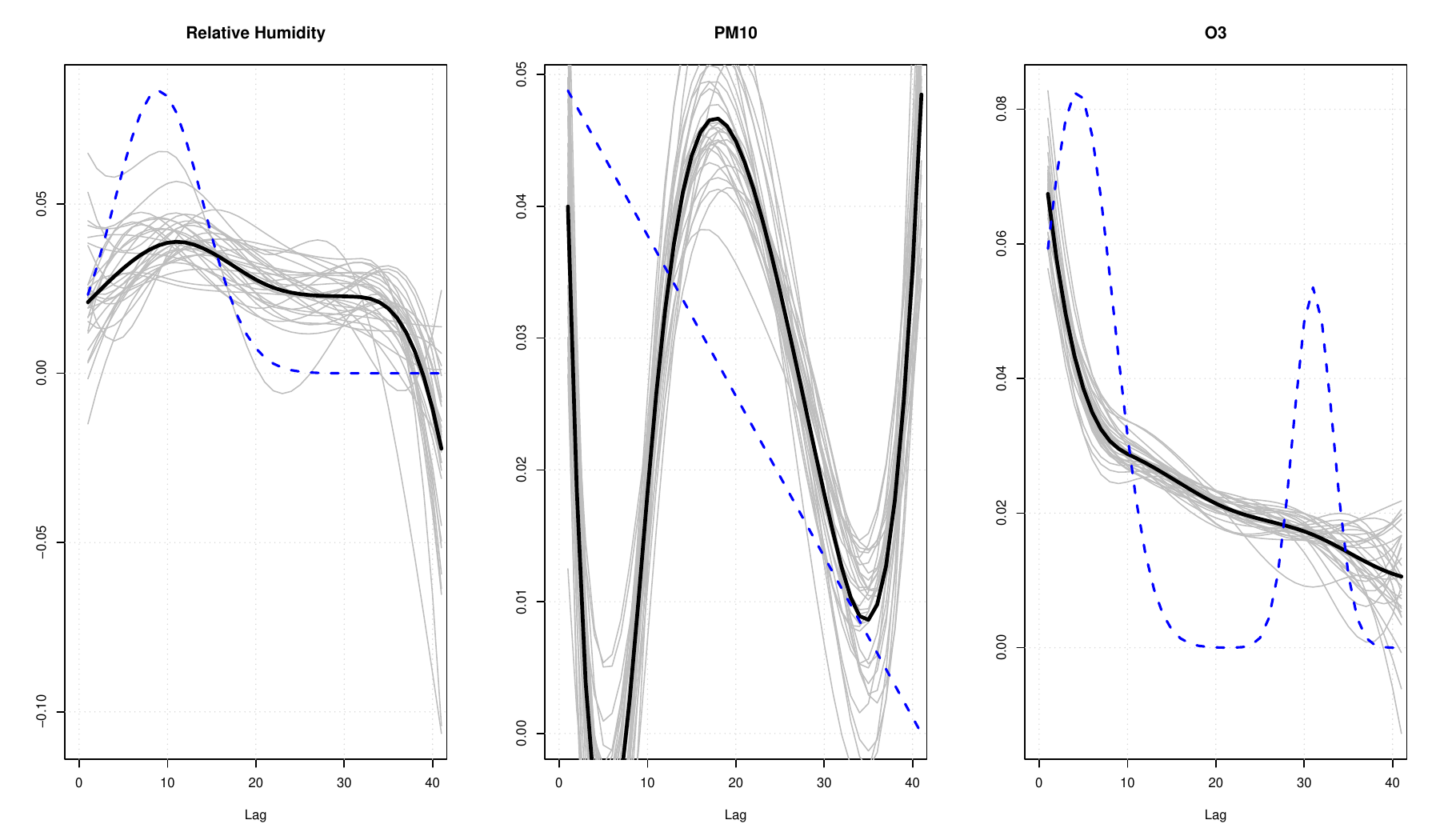}
	\end{subfigure}
	\captionof{figure}{Estimated lag-response for the first and second small simulations.}
\end{figure}

\setcounter{figure}{16}
\begin{figure}[hbt!]
	\centering
	\begin{subfigure}[b]{37em}
		\centering
		\includegraphics[scale=0.5]{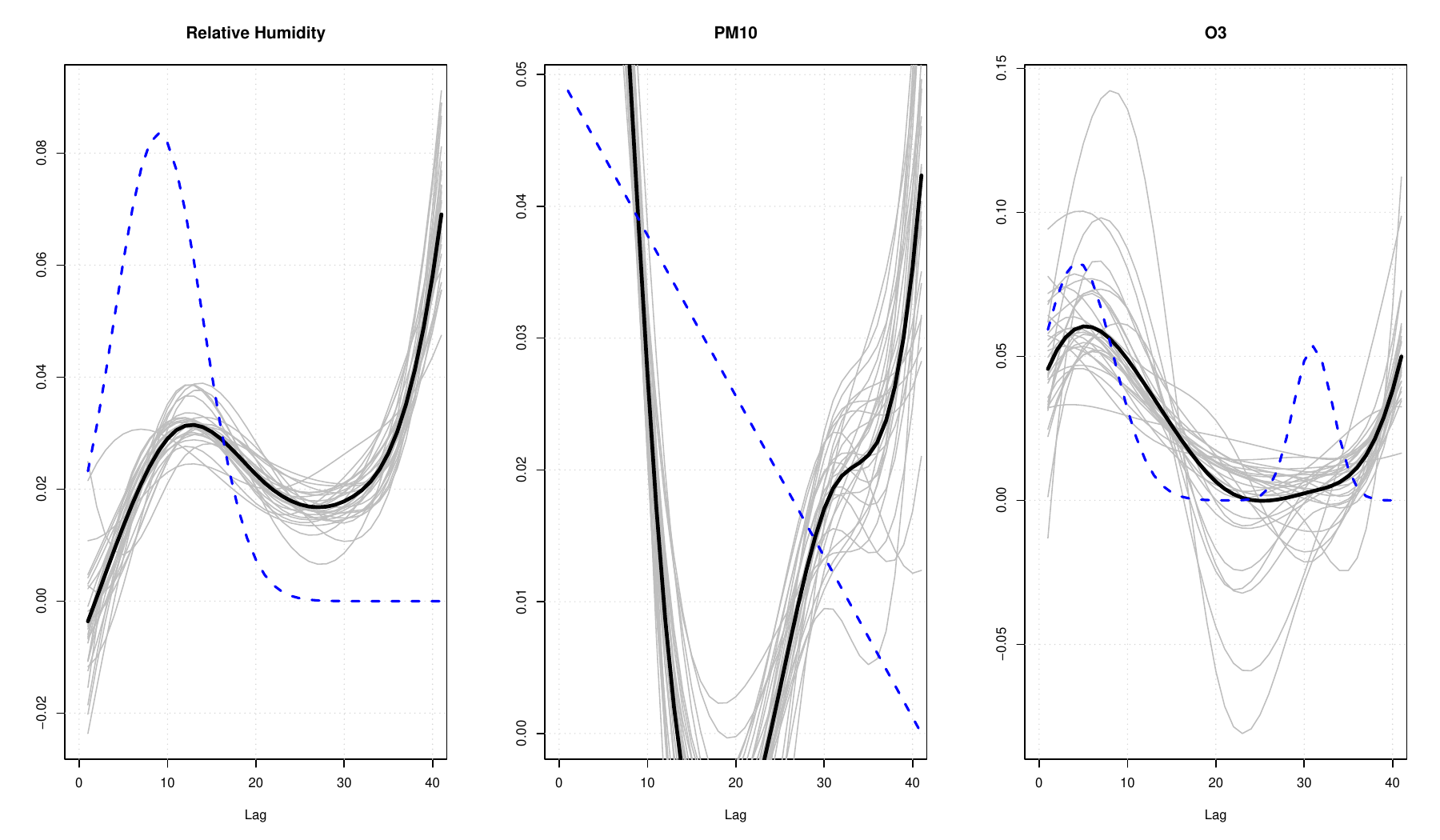}
	\end{subfigure}
	\hfill
	\begin{subfigure}[b]{37em}
		\centering
		\includegraphics[scale=0.5]{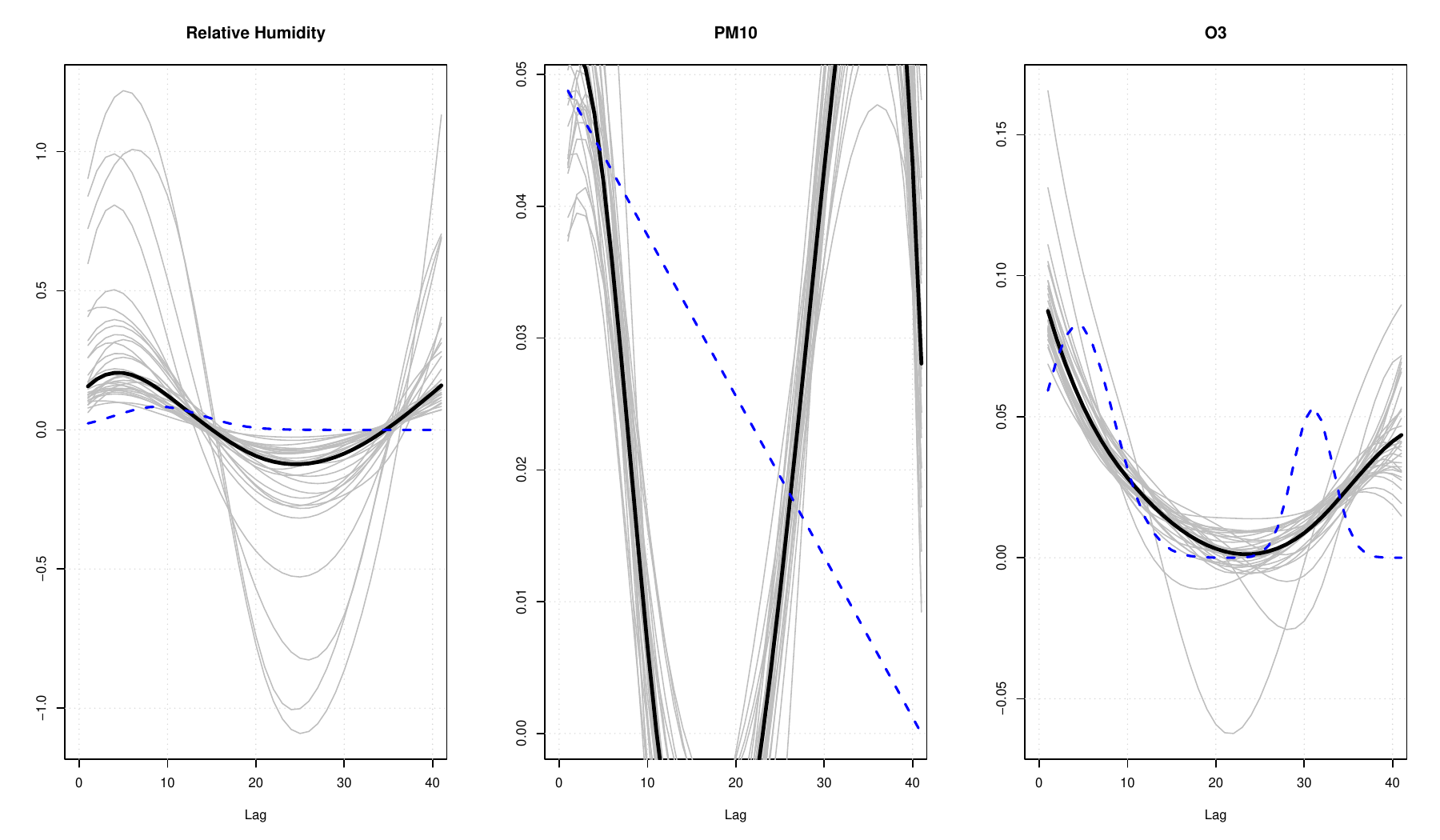}
	\end{subfigure}
	\captionof{figure}{Estimated lag-response for the third and fourth small simulations.}
\end{figure}
